%% file: paper.tex
\documentclass[sigconf,screen]{acmart}

\usepackage{flushend}
\usepackage{xspace}
\usepackage{tikz}
\usepackage{fancyvrb}
\usepackage{pifont}
\usepackage{cleveref}
\usepackage{subfig}

\usepackage{color}
\usepackage{tcolorbox}

\definecolor{lightgrey}{RGB}{244,244,244}
\definecolor{darkgrey}{RGB}{99,99,99}

\newcommand{\sys}{FlexOS\xspace}
\newcommand{\LibOS}{LibOS\xspace}

\newcommand*\BC[1]{\tikz[baseline=(char.base)]{
	\node[shape=circle,draw,inner sep=0.15pt] (char) {\textcolor{black}{#1}};}}

\copyrightyear{2022}
\acmYear{2022}
\setcopyright{acmcopyright}
\acmConference[ASPLOS '22]{Proceedings of the 27th ACM International Conference on Architectural Support for Programming Languages and Operating Systems}{February 28 -- March 4, 2022}{Lausanne, Switzerland}
\acmBooktitle{Proceedings of the 27th ACM International Conference on Architectural Support for Programming Languages and Operating Systems (ASPLOS '22), February 28 -- March 4, 2022, Lausanne, Switzerland}
\acmPrice{15.00}
\acmDOI{10.1145/3503222.3507759}
\acmISBN{978-1-4503-9205-1/22/02}

\begin{document}

\title{\sys: Towards Flexible OS Isolation}

\author{Hugo Lefeuvre}
\affiliation{
    \institution{The University of Manchester}
    \city{Manchester}
    \country{UK}
}

\author{Vlad-Andrei Bădoiu}
\affiliation{
    \institution{University Politehnica of Bucharest}
    \city{Bucharest}
    \country{Romania}
}

\author{Alexander Jung}
\affiliation{
    \institution{Lancaster University / \emph{Unikraft.io}}
    \city{Lancaster}
    \country{UK}
}

\author{Stefan Lucian Teodorescu}
\affiliation{
    \institution{University Politehnica of Bucharest}
    \city{Bucharest}
    \country{Romania}
}

\author{Sebastian Rauch}
\affiliation{
    \institution{Karlsruhe Institute of Technology}
    \city{Karlsruhe}
    \country{Germany}
}

\author{Felipe Huici}
\affiliation{
    \institution{NEC Labs Europe / \emph{Unikraft.io}}
    \city{Heidelberg}
    \country{Germany}
}

\author{Costin Raiciu}
\affiliation{
    \institution{UPB / \emph{Correct Networks}}
    \city{Bucharest}
    \country{Romania}
}
\authornote{UPB: University Politehnica of Bucharest}

\author{Pierre Olivier}
\affiliation{
    \institution{The University of Manchester}
    \city{Manchester}
    \country{UK}
}




\renewcommand{\shortauthors}{Lefeuvre et al.}

\input{00-abstract}

\date{}
\maketitle

\thispagestyle{empty}

\input{01-introduction}
\input{02-principles-challenges}
\input{03-design}

\input{04-implementation}

\input{05-exploration}
\input{06-evaluation}
\input{06bis-impact}
\input{07-related-works}
\input{08-conclusion}

\section*{Acknowledgements}

We would like to thank the anonymous reviewers, and our shepherd, Gerd
Zellweger, for their comments and insights. A similar thanks goes to our
colleague Marc Rittinghaus for his insights.  We are immensely grateful to the
Unikraft OSS community for their past and ongoing contributions.  This work was
funded by a studentship from NEC Labs Europe, EU H2020 grants 825377 (UNICORE),
871793 (ACCORDION) and 758815 (CORNET), as well as the UK’s EPSRC grants
EP/V012134/1 (UniFaaS) and EP/V000225/1 (SCorCH). UPB authors were partly supported
by VMWare gift funding.

\input{ae}

\bibliographystyle{ACM-Reference-Format}
\bibliography{bib}

\end{document}

%% file: 00-abstract.tex
\begin{abstract}
\noindent At design time, modern operating systems are locked in a specific
safety and isolation strategy that mixes one or more hardware/software
protection mechanisms (e.g. user/kernel separation); revisiting these choices
after deployment requires a major refactoring effort. This rigid approach shows
its limits given the wide variety of modern applications’ safety/performance
requirements, when new hardware isolation mechanisms are rolled out, or when
existing ones break.

We present \sys, a novel OS allowing users to easily specialize the safety and
isolation strategy of an OS at compilation/deployment time instead of design
time. This modular LibOS is composed of fine-grained components that can be
isolated via a range of hardware protection mechanisms with various data
sharing strategies and additional software hardening. The OS ships with an
exploration technique helping the user navigate the vast safety/performance
design space it unlocks. We implement a prototype of the system and
demonstrate, for several applications (Redis/Nginx/SQLite), \sys' vast
configuration space as well as the efficiency of the exploration technique: we
evaluate 80 \sys configurations for Redis and show how that space can be
probabilistically subset to the 5 safest ones under a given performance budget.
We also show that, under equivalent configurations, \sys performs similarly or
better than existing solutions which use fixed safety configurations.

\end{abstract}

%% file: 01-introduction.tex
\section{Introduction}
\label{sec:introduction}

\begin{figure}
    \includegraphics[width=\linewidth]{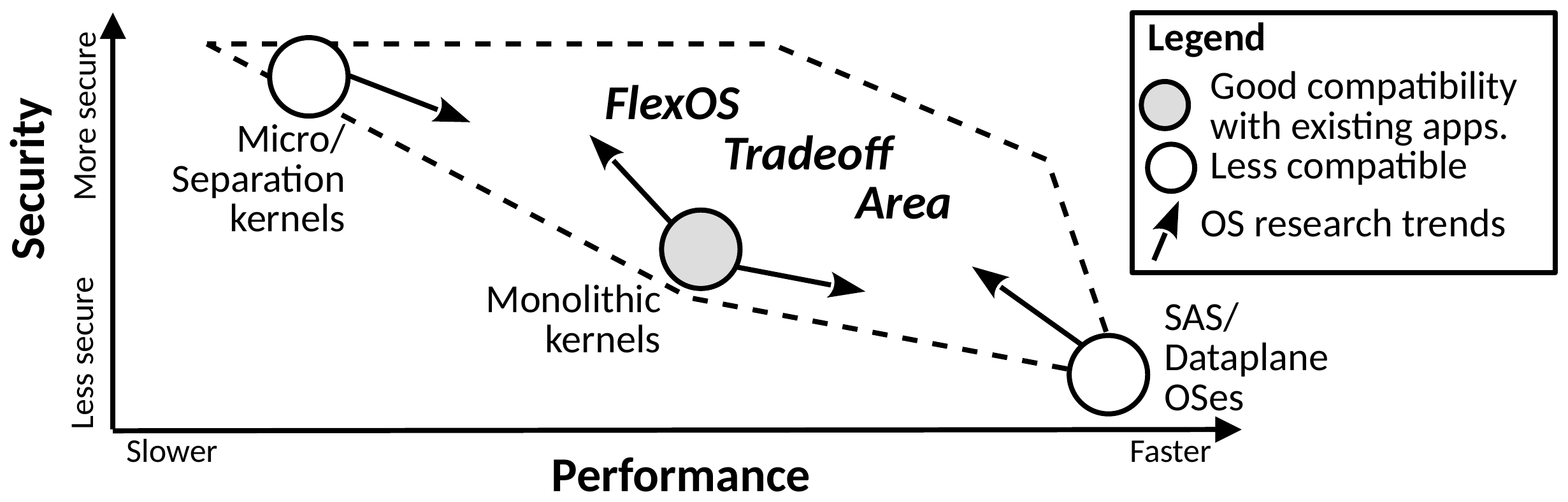}
    \caption{Design space of OS kernels.}
    \label{fig:design-space}
\end{figure}

Modern OS architectures are heavily interlinked with the protection mechanisms
they rely upon. OSes rigidly commit at design time to various high-level safety
decisions, such as the use of software verification, hardware isolation,
runtime checking, etc. Changing these after deployment is rare and costly.

The current OS design landscape (depicted in Figure~\ref{fig:design-space}) broadly consists of
micro-kernels~\cite{Herder2006, Klein2009}, which favor hardware
protection and verification over performance, monolithic
kernels~\cite{bovet2005}, which choose privilege separation and
multiple address spaces (ASes) to isolate applications, but assume all
kernel code is trusted, and single-address-space OSes (SASOSes), which
attempt to bring isolation within the address space~\cite{Chase1994,
  Leslie1996, Heiser1999}, or ditch all protection for maximum
performance~\cite{Madhavapeddy2013, Olivier2019, Kuenzer2021}. Making
post-design changes to these high-level safety decisions is very
difficult to implement. For instance, removing the user/kernel
separation~\cite{Maeda2003} requires a lot of engineering effort, as
does breaking down a process into multiple address spaces for
isolation~\cite{Kilpatrick2003}. Recently, the potential safety
benefits hinted by the proposal to introduce Rust components in
Linux~\cite{RUST_LINUX1} are questioned by the fact that the bulk of
the kernel code will remain written in a memory-unsafe
language~\cite{RUST_LINUX2}.

The rigid use of safety primitives in modern OSes poses a number of problems.
First, it precludes per-application OS specialization~\cite{Engler1995,
Kaashoek1997, Manco2017, Martins2014} at a time when modern applications
exhibit a wide range of safety and performance requirements. Prematurely
locking the design into any combination of safety primitives is likely
to result in suboptimal performance/safety in many scenarios. Effortless
specialization for safety is further motivated by the fact that today's
applications are made up of multiple components showing different degrees of
trust and criticality, and as such requiring various levels of isolation.
Furthermore, new isolation mechanisms~\cite{MPK, Schrammel2020, Watson2015,
Costan2016, ARMTrustZone2009, ARMMorello2020}, with the ability to complement
or replace traditional ones, are regularly being proposed by CPU manufacturers.
When multiple mechanisms can be used for the same task, choosing the most
suitable primitive depends on many factors, and should ideally be postponed to
deployment time. Finally, when the protection offered by a hardware primitive
breaks down (e.g. Meltdown~\cite{MELTDOWN}), it is difficult to decide how it should be
replaced, and generally costly to do so.

This leads us to the following research problem: \textit{how can we enable
users to easily and safely switch between different isolation and protection
primitives at deployment time, avoiding the lock-in that characterizes the
status-quo?}

%Most software, including OSes, integrates modules from different sources, with
%various levels of trust. Unfortunately, the properties assumed by the module
%designers affect the way in which a module can be used, limiting its
%usefulness. Take, for instance, a formally verified OS subsystem: how does one
%go about embedding it into a larger project while still maintaining its safety
%properties? Clearly, embedding it alongside untrusted C code may mean that its
%verified properties may not hold in practice. Thus, we ask \textit{how can we
%use the isolation and protection primitives such that the safety properties of
%a library will hold at runtime without isolating each library independently?}

%Lastly, computer hardware is becoming heterogeneous~\cite{SECPOPCORN_HET} and
%certain primitives are hardware-dependent (e.g. Intel Memory Protection Keys --
%MPK~\cite{MPK}, CHERI~\cite{Watson2015}). When running the same software on
%different hardware, how can we minimize the porting effort while preserving
%safety?

Our answer is \emph{\sys}, a modular OS design whose compartmentalization and
protection profile can easily and cost-efficiently be tailored towards a
specific application or use-case at build time, as opposed to design time. To
that aim, we extend the Library OS (\LibOS) model and augment its capacity to be
specialized towards a given use case, historically done for performance
reasons~\cite{Engler1995, Kaashoek1997, Manco2017, Martins2014}, towards the
\emph{safety} dimension.

With \sys, the user can decide, at \emph{build time}, which of the fine-grained
OS components should be placed in which compartment (e.g. the scheduler, TCP/IP stack,
etc.), how to instantiate isolation and protection primitives for each
compartment, what data sharing strategies to use for communication between
compartments, as well as what software hardening mechanisms should be applied
on which compartments. To that aim, we abstract the common operations required
when compartmentalizing arbitrary software behind a generic API that is used to
retrofit an existing \LibOS into \sys. This API limits the manual porting effort
of kernel and application legacy components to the marking of shared data using
annotations. These annotations, alongside other abstract source-level
constructs, are replaced at build time by a code transformation step that
instantiate a given \sys safety configuration.

The design space enabled by the system, illustrated on
Figure~\ref{fig:design-space}, is very large and difficult for a non-expert
user to explore manually. This leads to the second research question we
explore: \emph{how to guide the user navigating the vast design space unlocked
by \sys?} To answer this, we propose a semi-automated exploration
technique named \emph{partial safety ordering}, using partially ordered sets to
describe the probabilistic security degrees of \sys' configurations and
identify the safest ones under a given performance budget.

%% In summary, our research contributions are:
%% \begin{compactitem}
%% \item
%% The design of \sys, a novel OS whose safety/isolation profile can easily be
%% adapted to a wide range of technologies and granularities.

%% \item
%% A set of semi-automated design space exploration techniques helping the user to
%% select suitable \sys' configurations for a given use case.

%% \item
%% A prototype with support for Intel MPK and VM/EPT-based isolation.

%% \item
%% An evaluation of the system demonstrating over four scenarios the vast
%% safety/performance design it enables and the efficiency of the exploration
%% techniques.
%% \end{compactitem}

We have implemented a prototype of \sys with support for Intel MPK and
VM/EPT-based isolation, as well as a wide range of hardening mechanisms (CFI \cite{Abadi2009},
ASAN \cite{LinuxASAN}, etc.). Our evaluation using four popular applications demonstrates
the wide safety versus performance tradeoff space unlocked by \sys: we evaluate over
160 configurations for Redis and Nginx. We also show the ease of exploring
different points in that space: our semi-automated exploration technique can
probabilistically subset the 80 Redis configurations to the 5 safest ones
under a given performance budget. Finally, we demonstrate that under equivalent
configurations, \sys performs better or similarly to baselines/competitors: a
monolithic kernel, a SASOS, a microkernel, and a compartmentalized \LibOS.

%% techniques.

%% file: 02-principles-challenges.tex
\section{Flexible OS Isolation: Principles, Challenges}\label{sec:approach}

\sys seeks to enable users to easily and safely switch between different
isolation and protection primitives at deployment time. This section formalizes
the fundamental design principles required to achieve this, the challenges that
arise from them, and how we address them.

\subsection{Principles}

\noindent (P1) \textbf{\emph{The isolation granularity of \sys' components
should be configurable.}} The compartmentalization strategy, i.e. the number of
compartments and which components are merged/split into compartments, has a
major impact on safety and performance, thus it should be configurable.

\noindent (P2) \textbf{\emph{The hardware isolation mechanisms used should be
configurable.}} There is a wide range of isolation mechanisms with various
safety and performance implications. These should be configurable by the user.
For the OS developer, supporting a new mechanism should not involve any
rewrite/redesign and be as simple as implementing a well-defined API.

\noindent (P3) \textbf{\emph{Software hardening and isolation mechanisms should
be configurable.}} Software hardening techniques such as CFI, or Software Fault
Isolation (SFI), as well as memory safe languages such as Rust, bring different
levels safety at a variable performance cost. They should be selectively
applicable on the components they are the most meaningful for in a given use
case.

\noindent (P4) \textbf{\emph{Flexibility should not come at the cost of
performance.}} The OS runtime performance should be similar to what would be
achieved with any particular safety configuration without the flexibility
approach.

\noindent (P5) \textbf{\emph{Compatibility with existing software should not
come at a high porting cost}}, to maximize adoption.

\noindent (P6) \textbf{\emph{The user should be guided in the vast design space
enabled by \sys.}} Given its very large configuration space, the system should
come with tools helping the user identify suitable safety/performance
configurations for a given use case.

\subsection{Challenges and Approach}

P1 and P4 raise the question of \textbf{\emph{how to offer variable isolation
granularities, and how to do so without compromising performance?}} Genericity
is typically paid at the price of performance~\cite{Martins2014, Marinos2014,
Kuenzer2021}, and interface design may not be easily decoupled from the
isolation granularity without performance loss~\cite{Gefflaut2000}.
In order to tackle this issue, we propose to rely on a \LibOS
design that is \textit{already finely modularized while providing state of the
art performance}, Unikraft~\cite{Kuenzer2021}. The main idea is to consider
Unikraft's level of modularization (micro-library) as a minimal granularity,
using pre-existing interfaces as compartment boundaries. Then, in order to
maximize performance and safety for a given use case, less granular
configurations can be composed by merging select components into compartments.
At build time when an isolation mechanism is selected, \sys uses code
transformations to inline function-call-like cross-domain gates, avoiding the
overhead of a runtime abstraction interface~\cite{Ford1997}.

P2 and P5 bring the challenge of \textbf{\emph{how to design an OS in which 1)
isolation can be enforced by many hardware mechanisms and 2) the engineering
cost of introducing a new mechanism is low?}} Technology agnosticism is already
difficult in userland software, but core kernel facilities (interrupt handling,
memory management, scheduling) introduce additional complexity that should be
handled very differently depending on the underlying isolation technology. For
example, some technologies share a single address space between protection
domains (e.g. MPK~\cite{MPK}) while other use disjoint address spaces (e.g.
TEEs~\cite{ARMTrustZone2009}, EPT). The main idea of \sys is to abstract
existing isolation technologies and identify kernel facilities that
require different handling depending on the technology, and design these
subsystems so as to minimize the changes needed when implementing a new
technology.

P5 asks \textbf{\emph{how to limit the engineering costs of porting new
applications/libraries?}} To allow compatibility with existing software, \sys
extends an OS that offers a POSIX interface. That OS is compartmentalized by
marking cross-component calls and shared data using an abstract API and, in its
basic form, porting a new application requires the developer to use the same
API to mark shared data (i.e. data passed to other components) with
source-level annotations. This avoids the need to change the application design
or major code rewriting. Such an approach is common among state-of-the art
compartmentalization frameworks~\cite{VahldiekOberwagner2019, Hedayati2019,
Narayan2020, Schrammel2020}.

Finally, P1-P3 and P6 raise the question of \textbf{\emph{how to help the user
navigate the vast design space enabled by \sys?}} The introduction of safety
flexibility increases the potential for safety/performance specialization, but
selecting suitable configurations may be hard for a non-expert. For example, it
can be difficult to reason about the safety implications of increasing the
degree of compartmentalization vs. increasing the level of software hardening
for a given configuration. To tackle that issue, we propose a method named
\emph{partial safety ordering}, using partial order relationships to
probabilistically rank \sys configurations by safety and identify the
safest ones for a given application under a performance budget.

Section~\ref{sec:design} presents an OS design that satisfies P1-P5, and
Section~\ref{sec:implementation} gives key implementation points of a
prototype we developed. Section~\ref{sec:exploration} shows an approach to
tackle P6. Finally, Section~\ref{sec:evaluation}
presents an evaluation of our prototype.

%% file: 03-design.tex
\section{Designing an OS with Flexible Isolation}\label{sec:design}

We now provide an overview of \sys' main elements, starting with an overview 
of its design, compartmentalization API, the backend API, and finally the trusted computing
base.

%\subsection{Design Overview}

\begin{figure}
\center
\includegraphics[width=0.45\textwidth]{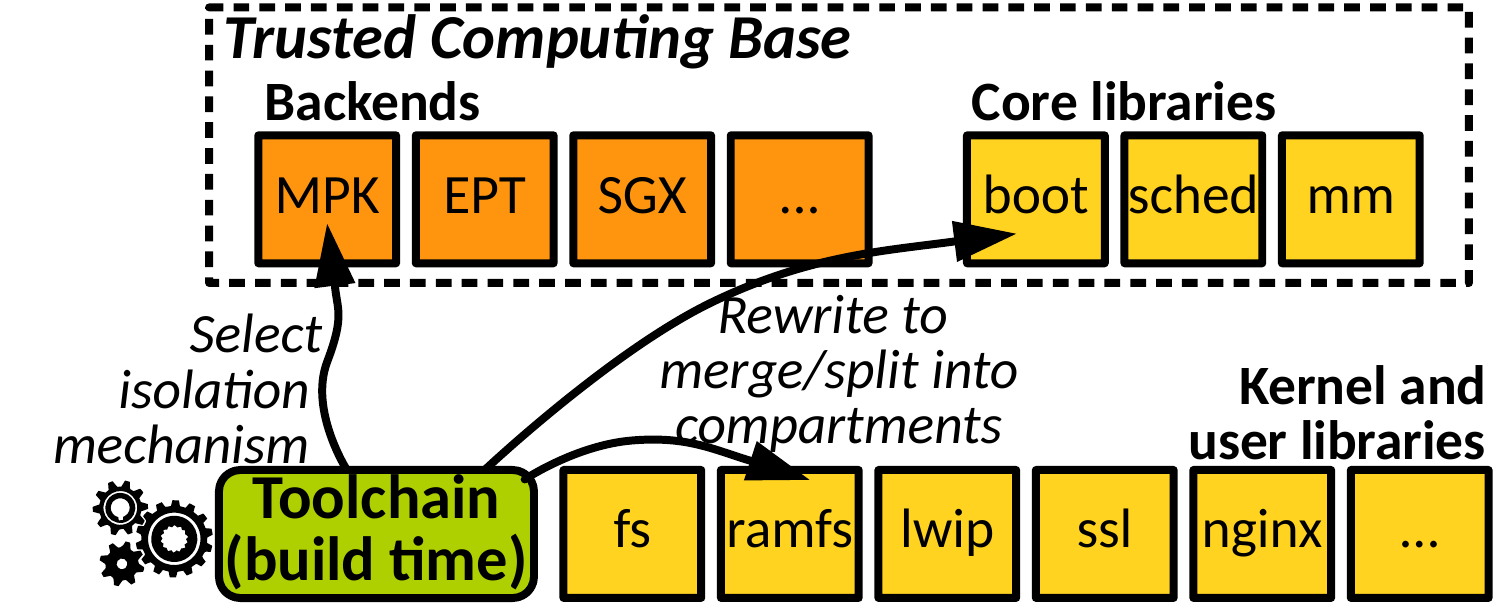}
\caption{
OS overview. The TCB includes backends and core libraries. Backends are used
by the toolchain to rewrite the libraries at build time.
}
\label{fig:shortoverview}
\end{figure}

\sys is based on a modular \LibOS, Unikraft~\cite{Kuenzer2021} composed
of a set of independent, fine grained libraries. 
In \sys, each library can be placed in a given
compartment (an isolation domain), and it can be hardened via techniques
such as Control-Flow Integrity (CFI), address sanitization and so forth. This safety
configuration is provided at build time, in a configuration file provided by
the developer, and \sys' toolchain produces an OS image with the desired safety
characteristics. Below is an example of such a configuration file that isolates
libopenjpg and lwip in a separate compartment with CFI and ASan enabled.

\begin{tcolorbox}[colback=lightgrey,boxrule=0pt,arc=0pt,left=6pt]
{\scriptsize
\begin{Verbatim}[commandchars=\\\{\}]
compartments:
  - comp1:
    mechanism: \textit{intel-mpk}
    default: True
  - comp2:
    mechanism: \textit{intel-mpk}
    hardening: [cfi, asan]
libraries:
  - libredis: comp1
  - libopenjpg: comp2
  - lwip: comp2
\end{Verbatim}
}
\end{tcolorbox}

In contrast to Unikraft where all libraries are in the same protection domain
and any library can directly call a function from another
library, in \sys' source code libraries call external functions via
\textit{abstract gates}, and may share data with external libraries at the granularity of a
byte using abstract code annotations. Gates and annotations form an API used to
compartmentalize Unikraft into \sys, and represent metadata which is
automatically replaced by our toolchain with a particular implementation
at build time.  Different implementations can leverage different isolation
technologies, or flavors of a same technology. We refer to the API
implementation for a given technology (MPK, EPT, etc.) together with its
runtime library as \textit{isolation backend}. This subsection gives a short
overview of \sys' main design elements, which are then elaborated in the
following subsections. Figure~\ref{fig:shortoverview} depicts the components
described in this subsection.

\paragraph{\LibOS Basis.} Achieving flexible isolation at a fine granularity
implies a high degree of modularity. In practice, this modularity is not
offered by typical monolithic general-purpose OSes~\cite{Kuenzer2021}. A
flexible isolation approach on the basis of Linux would require a first
non-trivial ``modularization'' step~\cite{Li2021} that may take years of
engineering and careful redesign. Library OSes~\cite{Kuenzer2021} and
component-based OSes~\cite{Bruno1999, Parmer2007} are a better starting point
for flexible OS isolation because they often provide highly modular code bases
with good application compatibility and high performance. Flexible isolation
also suits well the specialization spirit of \LibOS{}es, where the OS can be
tailored for a given application/use-case. This was historically done for
performance~\cite{Engler1995}, and \sys enables specialization towards safety.

\paragraph{API and Build-time Instantiation.} Unlike a typical \LibOS, we design
\sys in an \textit{isolation-agnostic} manner. Cross compartment calls are made
through abstract call gates that are instantiated at build time (arrows in
Figure~\ref{fig:shortoverview}). Shared data is marked using compiler
annotations, used at build time to instantiate a given data sharing strategy.
Unlike linker-based approaches~\cite{Sartakov2021}, \sys performs
replacements using source to source transformations using
Coccinelle~\cite{COCCINELLE1, COCCINELLE2}.  This has the advantage of allowing
all compiler optimizations and gives \sys a clear performance advantage
compared to historical approaches that relied on heavyweight runtime
abstraction interfaces such as \texttt{COM} for Flux OSKit~\cite{Ford1997}. It
also makes \sys' isolation approach easy to debug and understand by anyone who
knows C: transformations can be visually inspected in a high-level language
with usual file comparison tools.

% Hugo: can we make use of this? Is it a good idea?
%\begin{table}[]
%\begin{tabular}{|l|c|c|c|c|c|}
%\hline
%Properties & \multicolumn{1}{l|}{MPK} & \multicolumn{1}{l|}{\parbox[t]{0.5cm}{EPT/\\RPC}} &
%             \multicolumn{1}{l|}{SGX} & \multicolumn{1}{l|}{\parbox[t]{0.5cm}{Trust\\Zone}} &
%             \multicolumn{1}{l|}{CHERI} \\ \hline \hline
%Granularity                               & page & page & page & page & byte \\ \hline
%\parbox[t]{2.5cm}{Arbitrary\\READ Prot.}  & \ding{51} & \ding{51} & \ding{55} & \ding{55} & \ding{51} \\ \hline
%\parbox[t]{2.5cm}{Arbitrary\\WRITE Prot.} & \ding{51} & \ding{51} & \ding{55} & \ding{55} & \ding{51} \\ \hline
%\parbox[t]{2.5cm}{Arbitrary\\EXEC Prot.}  & \ding{55} & \ding{51} & \ding{55} & \ding{55} & \ding{51} \\ \hline
%Monotonicity                              & \ding{55} & \ding{55} & \ding{55} & \ding{55} & \ding{51} \\ \hline
%\parbox[t]{2.5cm}{Maximum\\No. of Domains}& 16   & -    & -    & 1    & -    \\ \hline
%\end{tabular}
%\caption{Properties of different hardware isolation mechanisms.}
%\label{tab:hwmechanisms}
%\end{table}

\begin{figure*}
    \center
    \includegraphics[width=\textwidth]{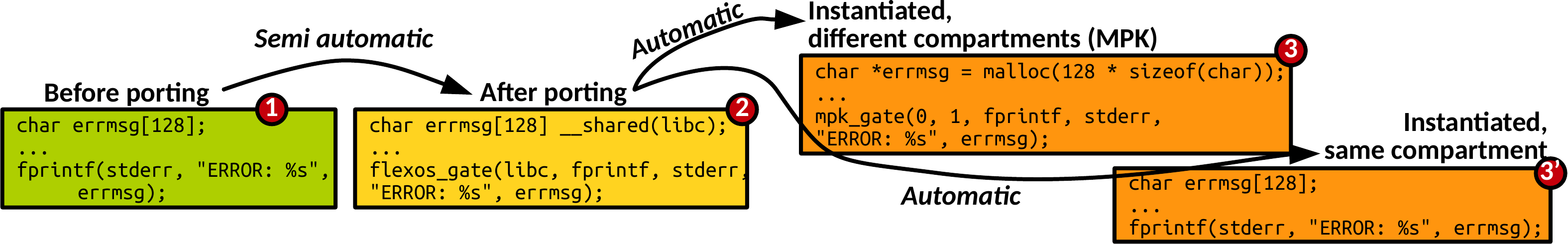}
    \caption{ \sys code transformations. First, developers manually annotate
        shared data, and gate placeholders are automatically inserted. At build
        time, API primitives are automatically replaced with the chosen
        mechanism. In the MPK case, shared data can for example be allocated
    on a shared heap. If the two libraries are in the same compartment, the
result is similar to the code prior porting, resulting in zero overhead. }

    \label{fig:porting}
\end{figure*}

\subsection{Compartmentalization API and Transformations}\label{subsec:api}

Most isolation mechanisms (memory protection keys~\cite{MPK}, TEEs like
SGX~\cite{Costan2016}, or hardware capabilities~\cite{Watson2015}) restrict data
access according to a set of current privileges,
%(typically defined by a register) within a single system.
and provide a means to switch
privileges %(typically a specific instruction)
and share data across compartments.
%(typically shared memory or capabilities).
Ensuring safety is equivalent to controlling privilege transitions, making sure
that the system only ever enters ``legal'' couplings of executing code and data
privileges.  Other isolation approaches such as ARM
TrustZone~\cite{ARMTrustZone2009} or EPT/VMs consider compartments as entirely
different systems (or ``worlds''), enforcing a 1:1 system/compartment mapping.
With this approach, systems never switch privileges, instead they communicate
with other compartments via remote procedure calls (RPCs) and shared memory.
We design \sys' call gates and data sharing primitives to cater for both
approaches.  In \sys, the only requirement for an isolation mechanism is to (1)
implement the concept of protection domains and provide a domain switching
mechanism, and (2) support some form of shared memory for cross-domains
communication. To the best of our knowledge, this applies to the vast majority
of industry and research isolation mechanisms. This subsection gives an
overview of \sys' compartmentalization approach, first focusing on the API with
call gates and shared data, and then on build-time source transformations.

\paragraph{Call Gates.}

In \sys, cross-library calls are represented in the source code by
\textit{abstract call gates}. At build time, as part of the transformation
phase, abstract call gates are replaced with a specific implementation. For
instance, when the caller and callee are configured to be in the same
compartment, call gates implement a classical function call. When they are in
different compartments, isolated for example by MPK, the call gate performs a
protection domain switch before finally executing the \texttt{call}
instruction. In a setting where libraries are isolated using VMs, the call gate
performs a remote procedure call (RPC). From the perspective of the compiler,
caller, and the callee, call gates are entirely transparent as they implement
the System V ABI calling convention.  Unlike typical System V function calls
however, call gates guarantee isolation of the register set and therefore save
and zero out all registers not used by parameters. Figure~\ref{fig:porting}
presents an example of gates from the porting (step \BC{2}) to the replacement
by the toolchain (\BC{3} and \BC{3'}).

The part of the process of porting existing user/kernel code to \sys consisting
in marking call gates is automated: knowing the control-flow graph of the
system, static analysis determines whether a procedure call crosses library
boundaries, and if so, performs a syntactic replacement of the function call
with a call gate instead. A corner case requiring programming effort is when a
component calls another component through a function pointer. The callee cannot
be determined statically, thus the programmer must annotate the possible
pointed functions with the list of possible components they can be called from.
The toolchain will then generate wrappers enclosing the implementations of the
functions in question in the appropriate call gates. Our prototype
implementation uses Cscope~\cite{CSCOPE} and Coccinelle~\cite{COCCINELLE1}.

\sys call gates are not trampolines. Instead, they replace System V function
calls entirely and are always inlined at the call site. An advantage of such
approach is that call gates naturally provide an inexpensive (albeit
incomplete) form of CFI, guaranteeing that libraries can only be entered
through well defined entry points, known and enforced at compile time.

%One of the implications of hardcoding gates at build time is that the isolation
%profile of the image is fixed at runtime. This is a deliberate choice: the
%ability to create and destroy domains at runtime (or change a domain's
%permissions) implies an increased complexity in the TCB and less potential for
%static checks in the gates. Further, this is not a limitating choice: because
%\sys boots extremely fast (3ms, similar to a \texttt{fork/exec}), it is simple
%to reboot with a different image to achieve another isolation profile.

\paragraph{Data Ownership Approach.}

\sys takes a code-centered~\cite{Gudka2015} isolation approach. Each
library is present only once and maps to a specific set of privileges. There
is a slight tweak for backends that rely on several systems
(TrustZone, VMs): for them, the trusted computing base (\S\ref{sec:tcb})
is duplicated; one for each
system, as each compartment must possess a self-contained kernel (\S\ref{subsec:ept}).

\sys considers all static and dynamic data allocated by a library as private by
default. Individual variables can then be annotated as ``shared'' with a
specific group of libraries into \emph{whitelists}, similarly to access control
lists.  In practice, the maximum number of isolated data sharing ``zones'' is
limited by the underlying technology. Annotations are made with the keyword
\texttt{\_\_shared} as illustrated in Figure~\ref{fig:porting} step \BC{2}.

%Annotations look like this (a more complete example is depicted in
%Figure~\ref{fig:porting}):

%\begin{tcolorbox}[colback=lightgrey,boxrule=0pt,arc=0pt,left=6pt]
%{\footnotesize
 %\begin{Verbatim}[commandchars=\\\{\}, fontsize=\tiny]
%// annotation: __shared( [whitelist] )
%struct sockaddr_in addr; // stack, not shared
%struct sockaddr_in addr __shared(lwip); // stack, shared w/ lwip
 %\end{Verbatim}
%}
%\end{tcolorbox}

Compiler annotations are identical for all types of variables. However, under
the hood, \sys differentiates between statically allocated variables,
dynamically allocated heap variables, and dynamically allocated stack
variables.

\sys' compartmentalization API itself does not dictate \textit{how} variables
have to be shared. Different mechanisms can require very different sharing
approaches: while certain mechanisms such as MPK require shared data to be
located in shared memory regions, others such as CHERI's hybrid
capabilities~\cite{cheri} require compiler annotations that can be
automatically generated in place of the \sys placeholder.
Section~\ref{sec:implementation} describes the implementation of the API for
the two supported backends (MPK/EPT), and sketches implementations for an
additional one (CHERI).

Identifying shared data represents the vast majority of the porting effort. It
is necessary for both kernel libraries, user libraries, and applications. On
the kernel side, this problem is simplified (but not eliminated) by the modularity of Unikraft's
code base. This issue is not specific to \sys and is
widely explored in the literature. State of the art approaches (1) rely on
manual code annotations~\cite{Narayan2020}, (2) perform static analysis at
compile time to identify shared data automatically~\cite{Bauer2021}, or (3)
perform a mix of static, dynamic, and manual analysis~\cite{Gudka2015}.  There
is no silver bullet: manual code annotation can be non-trivial, but typically
produces precise results that not only take into account what \textit{is}
accessed across modules, but also what \textit{should be} shared from a
security perspective. Static-analysis based approaches, on the other hand, are
automatic, but conservative. These methods would be applicable to \sys, however
automated shared data identification is not the main focus of this paper. The
current prototype relies on manual annotations, and
Section~\ref{sec:implementation} details the porting effort for a number of
applications and libraries.

\paragraph{Build-time Source Transformations.}

Before compilation, \sys' toolchain performs source transformations to (1)
instantiate abstract gates, (2) instantiate data sharing code, (3) generate
linker scripts, and (4) generate additional code in core libraries according to
backend-provided recipes.  The amount code generated in considerable. As an
example, the toolchain modifies about 1 KLoC for a simple Redis configuration.
Figure~\ref{fig:porting} steps \BC{3} and \BC{3'} presents an example of the
porting-transformation process.

%As an example, the following simple semantic patch
%template used to replace gates took us more than 100 SLOCs with a Clang tool.

%\begin{tcolorbox}[colback=lightgrey,boxrule=0pt,arc=0pt,left=6pt]
%{\footnotesize
 %\begin{Verbatim}[commandchars=\\\{\}, fontsize=\tiny]
%@gatereplacer_noreturn@
%expression list EL;
%expression lname;
%@@
%- chrys_gate({{ lib_dest_name }}, EL);
%+ {{ gate }}({{ comp_cur_nb }}, {{ comp_dest_nb }}, EL);
 %\end{Verbatim}
%}
%\end{tcolorbox}

\subsection{Kernel Backend API}

Most isolation mechanisms require changes to a specific set of components in
the kernel. The kernel facilities that can require special handling
depending on the technology exclusively correspond to the core libraries (see
Figure~\ref{fig:shortoverview}). In order to make such changes scalable, we
designed core components to expose a \textit{hook API} to isolation backends,
allowing the core libraries to be easily extended with backend specific
functionalities. For example, the MPK backend leverages the thread creation
hook offered by the scheduler to switch a newly created thread to the
right protection domain. These hooks come at no cost: since the instantiation
is done at build time, the compiler is able to aggressively inline such calls.

Porting \sys to use a new isolation mechanism does not require redesign. In
general, it is equivalent to (1) implementing gates for the particular
mechanism, (2) implementing hooks for core components (see previous paragraph),
(3) implementing linker script generation in the toolchain, (4) implementing
Coccinelle code transformations, and (5) registering the newly created backend
into the toolchain. In practice, developers can heavily reuse existing
transformations for new backends.

\subsection{Trusted Computing Base}
\label{sec:tcb}

Regardless of the isolation mechanism, certain components are so deeply
involved in the OS' functioning that they will cause the entire system
to violate its safety guarantees when compromised.
These components are (1) the early boot code, (2) the memory manager, (3) the
scheduler, (4) the first-level interrupt handler's context switch primitives,
and (5) the isolation backend.  We refer to these components as \sys' trusted
computing base (TCB), illustrated in Figure~\ref{fig:shortoverview}.  Clearly,
malfunctioning or malicious early boot code can setup the system in an unsafe
manner, the memory manager can manipulate page table mappings in order to
freely access any compartment's memory, the scheduler can manipulate sleeping
thread's register states, and the backend provide incomplete isolation, etc.
This is the case even when considering architectural hardware capabilities such
as CHERI~\cite{Davis2019}.  It comes as no surprise: this ``core'' set of
libraries is historically the set of services that microkernel OSes
provide~\cite{Tanenbaum2006}.  \sys' TCB is small: around 3000 LoC in the case
of Intel MPK, and even less for VM/EPT.

\paragraph{Trust Model.}

The whole point of flexible isolation is to be able to achieve a wide range of
trust models where different components (such as the network stack, parser
libraries, etc.) can be considered untrusted and potentially compromised. Thus
there is no single trust model for \sys. In general, however, we assume that
the TCB (see previous paragraph) is safe and error free.  This is not an
unreasonable assumption given the small size and the potential for formal
verification (we have formally verified a version of our
scheduler~\cite{Lefeuvre2021} using Dafny~\cite{Leino2010}). The hardware and
the compiler are also part of the TCB.  Note that the rest of the toolchain
(Coccinelle included) is \textit{not} part of the TCB as the code includes
compile time checks that are able to detect invalid transformations. Finally we
must also assume that interfaces correctly check arguments and are free of
confused deputy/Iago~\cite{Checkoway2013} situations. This is not an
unreasonable assumption within the core \sys codebase. Further, confused deputy
and Iago attacks are probabilistically made more complex to execute in \sys due
to the variability of the interface size; the system call API, for example, is
divided into a variable number of sub-interfaces depending on the chosen
configuration, and several compartments may need to be subverted for an attack
to be successful.

%% file: 04-implementation.tex
\section{Prototype}\label{sec:implementation}

We present a prototype of \sys on top of Unikraft~\cite{Kuenzer2021} v0.5, with
Intel MPK and EPT backends. Modification to the Unikraft kernel represent about
3250 LoC: 1400 for the MPK backend, 1000 for EPT, and 850 for core libraries.
In user space, changes to Unikraft's toolchain represents 2300 LoC.  We port
user codebases (Redis, Nginx, iperf, and SQLite) as well as most kernel
components (the TCP/IP stack, scheduler, filesystem, etc.) to run as isolated
components.  This section presents the MPK and EPT backends, sketches a CHERI
backend, and concludes with the porting effort.

\subsection{Intel MPK Isolation Backend}

MPK is a mechanism present in Intel CPUs offering low-overhead intra-AS
memory isolation~\cite{inteldoc, Bannister2019, Schrammel2020}. MPK
leverages unused bits in the page table entries to store a \textit{memory
protection key}, enabling up to 16 protection domains. The PKRU register then
stores the protection key permissions for the current thread. On each memory
access, the MMU compares the key of the target page with the PKRU and triggers
a page-fault in case of insufficient permissions. \sys associates each
compartment with a protection key and reserves one key for a shared domain used
for communications. If the image features less than 15 compartments, \sys uses
remaining keys for additional shared domains between restricted groups of
compartments. Any compartment can modify the value of the PKRU, thus the MPK
backend has to prevent unauthorized writes. This has previously been done via
runtime checks~\cite{Hedayati2019} and static
analysis~\cite{VahldiekOberwagner2019}. In \sys, no code is loaded after
compilation, hence static binary analysis coupled with strict
\texttt{W$\oplus$X} is sufficient.

\paragraph{MPK Gates.}

For flexibility, \sys offers two different implementations of the MPK
gate. The main one provides full spatial safety, similarly
to HODOR~\cite{Hedayati2019}. The gate protects the register set and uses one
call stack per thread per compartment. Each compartment has a stack registry
that maps threads to their local compartment stack, making it fast and safe to
switch the call stack. Upon domain transition, the gate (1) saves the current domain's
registers set, (2) clears registers, and (3) loads function arguments. It then
(4) saves the current stack pointer, (5) switches thread
permissions, (6) switches the stack, and finally (7) executes the \texttt{call}
instruction. Once the function has returned, operations are executed in reverse.

The second gate implementation shares the stack and the register set across
compartments, similarly to ERIM~\cite{VahldiekOberwagner2019}. It is
conceptually very simple, switching the content of the PKRU before performing a
normal function call. This lightweight implementation offers lesser guarantees
but presents a lighter overhead, close to the raw cost of \texttt{wrpkru}
instructions.

\paragraph{Data Ownership.}

\sys' MPK images feature one data, read-only data, and bss section per
compartment to store private compartment static data.
%At build time, the
%toolchain maps \texttt{\_\_shared()} annotations on statically allocated
%variables to \texttt{section()} compiler attributes such that the compiler and
%linker automatically insert variables into the right section,  depending on the
%library's assigned compartment.
At boot time, the boot code protects
these sections with the compartment's protection key.

Each compartment has a private heap, and a shared one is used for
communications. Our prototype uses a single shared heap for all shared
allocations, but this is not a fundamental restriction.  Stack allocations are
slightly more complex. Existing works convert shared stack
allocations to shared heap allocations~\cite{Hedayati2019, Kjellqvist2020, Bauer2021}.
%The variable is then freed at all exit points of the function. While
%functionally correct,
This approach is costly from a performance perspective: an allocation+free on
the fast path for a modern allocator typically takes 30-60 cycles, and
up to thousands of cycles on the slow path~\cite{Kanev2017}. This is as
expensive as entire domain transitions, and that for a single shared
stack variable.  While \sys supports stack-to-heap conversions, we propose
another approach that addresses this issue, the \textit{data shadow stack}.

\paragraph{Data Shadow Stacks.}

\begin{figure}
    \center
    \includegraphics[width=0.43\textwidth]{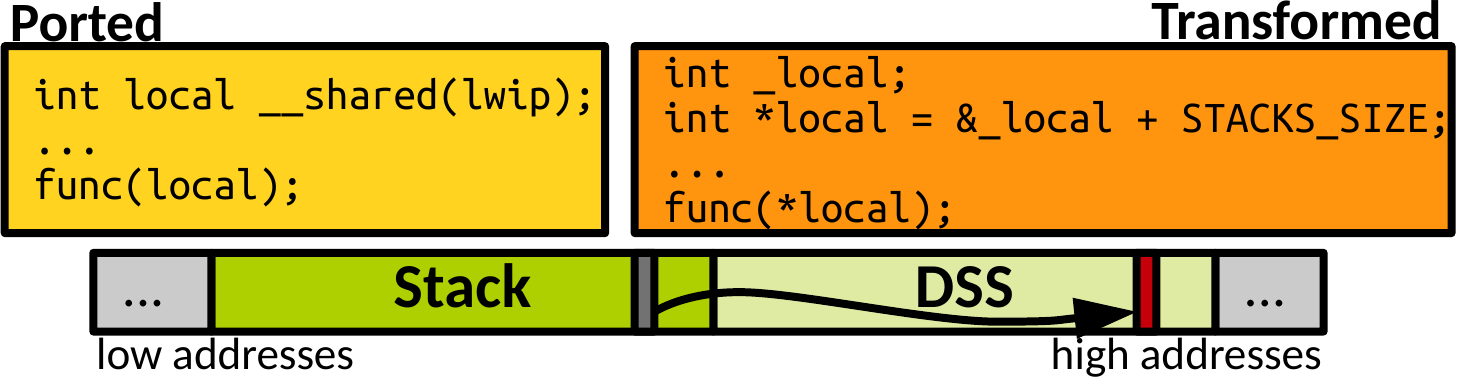}
    \caption{
Data Shadow Stacks.
}
    \label{fig:dss_fig}
\end{figure}

Stack allocations are much faster than heap allocations because the compiler is
able to perform bookkeeping \textit{at compile time}. At runtime, a single push
instruction is needed, resulting in constant low cost.  Data Shadow Stacks
(DSS), illustrated in Figure~\ref{fig:dss_fig}, leverage this bookkeeping work
for shared stack allocations.

When using the DSS, the usual stack size of threads is doubled. The upper part
corresponds to the DSS and is put in the shared domain.  The lower part is the
traditional stack and remains in the compartment's private domain. For each
shared variable \texttt{x}, we define the \textit{shadow} of \texttt{x} as \texttt{\&x
+ STACK\_SIZE}. Thus, allocating space for a shared variable on the stack
transparently allocates a shadow variable in the DSS.  Before compilation, the
toolchain replaces every reference to a shared stack variable with its shadow
\texttt{*(\&var + STACK\_SIZE)} in the shared domain.

Allocations on the DSS are much faster than on a shared heap, since the DSS'
bookkeeping overhead is null (stack speed), and the locality of reference high.
The cost is a relatively small increase in memory usage (stacks are twice as
large). The DSS mechanism is applicable to any isolation mechanism that
supports shared memory, and is compatible with common stack protection
mechanisms.
%We evaluate this mechanism in greater detail in
%\ref{subsec:microbench}.

%Note that the DSS does not fundamentally have to be next to the normal stack
%and might be placed anywhere in the address space --- at the expense of memory
%locality and possibly a bit of performance. This makes it applicable to other
%backends such as EPT. There might also be multiple DSSs to support several
%shared domains with different groups of compartments as discussed in
%\ref{subsec:api}.

\paragraph{Control Flow Integrity.}

Intel MPK does not provide protection from execution. As such, if a compartment
is compromised and the attacker ROPs into another compartment, a fault will not
directly happen. The MPK backend is able to provide a certain form of CFI,
ensuring that compartments can only be entered at well-defined points. This
ability is the consequence of the hardcoding of gates as described in
\ref{subsec:api}. If the control-flow of one compartment is compromised and the
attacker ROPs directly into another compartment $c$, then the system is
guaranteed to crash if any data local to $c$ is accessed.

\subsection{EPT/VM Backend}
\label{subsec:ept}

Virtualization has been used in many works to support isolation within a
kernel~\cite{LeVasseur2004, Nikolaev2013, Zhang2018, Nikolaev2020}.
Hardware-assisted virtualization is widely supported and provides strong safety
guarantees compared to MPK, at the cost of higher overheads. The EPT backend is
an extreme case; compartments do not share ASes and run on different vCPUs. It
shows that \sys is able to cater very different mechanisms under a common API.

\sys' EPT backend generates one VM image per compartment, each containing the
TCB (boot code, scheduler, memory manager, backend runtime) and the
compartment's libraries. Communications use a shared memory-based RPC
implementation. Our prototype runs on QEMU/KVM patched to support lightweight
inter-VM shared memory (less than 90 LoC).

\paragraph{EPT Gates.}

Upon domain transition, the caller places a function pointer and arguments in a
predefined shared area of memory. All other VMs busy wait until they notice an
RPC request, check that the function is a legal API entry point, execute the
function and place the return value in a predefined area of the shared memory.
In order to support multithreaded loads, each RPC server maintains a pool of
threads that are used to service RPCs.  Using function pointers instead of
abstract routine identifiers simplifies the RPC server's unmarshalling
operation and does not prevent the RPC server from checking the pointer to
ensure that it is a legal entry point. This optimization is possible since all
compartments are built at the same time, hence all addresses are known.

Busy-waiting allows the EPT backend to minimize gate latency as opposed to VM
notifications, but a similar implementation with \texttt{MONITOR/MWAIT}
instructions would also be possible to minimize power consumption if calls are
sparse.  Overall, any of these tweaks can be implemented as gate variant in
order to offer as much freedom as possible to the user.

\paragraph{Data Ownership.}

The EPT backend relies on shared memory areas to share data (static and
dynamic) across VMs. Areas are always mapped at the same address in the
different compartments so that pointers to/in shared structures remain valid.
Each VM manages its own portion of the shared memory area to avoid the need for
complex multithreaded bookkeeping.

\paragraph{Control Flow Integrity.}

The EPT backend is able to provide a form of CFI stronger than that of the MPK
backend, ensuring that compartments can only be \textit{left and entered} at
well defined points. Indeed, the RPC server can control at entry that the
executed function is legal, and compartments are not able to execute other
compartments' code without RPC calls.

\subsection{Supporting More Isolation Mechanisms}

To check whether \sys can support other isolation backends, we discuss how we
can leverage the CHERI hardware capabilities~\cite{Watson2015}, an emerging
isolation hardware mechanism. The CHERI ISA extension is available for ARMv8-A,
which is supported by \sys. Among others, CHERI capabilities would extend
FlexOS’ trade-off space with the ability to address confused-deputy
situations, reduce data sharing, and allow for a larger number of domains,
something that is currently impossible for architectural (MPK) and performance
(EPT) reasons.  The backend would use boot-time hooks to initialize CHERI
support, and scheduler hooks to perform capability-aware context-switching and
thread initialization.  Similarly to other backends, CHERI gates would save
caller context, clear the relevant traditional and capability registers,
install the callee context, and rely on the domain crossing instruction
\texttt{CInvoke} and sentry capabilities~\cite{Watson2020b} to perform
protection domain jumps. As a first step, \sys should rely on the hybrid
pointer model to maximize compatibility.  Our API's shared data annotations
would transform to \texttt{\_\_capability} at build time to treat shared
variables as a capabilities for efficient communications.
% In the long run, a pure-capability approach should be adopted
% for \sys core libraries, potentially with hybrid pointers for user-code (in
% order to avoid potential porting effort).

\begin{table}[]
\caption{Porting effort: size of the patch (including automatic gate
replacements), number of shared variables.}
\center\small
\label{tab:porting}
\begin{tabular}{l|l|l}
\hline
\textsc{Libs/Apps}                           & \textsc{Patch size} & \textsc{Shared vars} \\ \hline \hline
TCP/IP stack (LwIP)                          & +542 / -275        & 23                   \\ \hline
scheduler (\texttt{uksched})                 & +48 / -8           & 5                    \\ \hline
filesystem (\texttt{ramfs}, \texttt{vfscore})& +148 / -37         & 12                   \\ \hline
time subsystem (\texttt{uktime})             & +10 / -9           & 0                    \\ \hline
Redis                                        & +279 / -90         & 16                   \\ \hline
Nginx                                        & +470 / -85         & 36                   \\ \hline
SQLite                                       & +199 / -145        & 24                   \\ \hline
iPerf                                        & +15 / -14          & 4                    \\ \hline
\end{tabular}
\end{table}

\subsection{Porting Effort}

The porting process consists of two phases: call gate insertion (automated),
and shared data annotation (manual). The typical workflow, once gates have been
inserted, is to run the program with a representative test case (e.g., a
benchmark or test suite) until it crashes due to memory access violations.
Crash reports point to the symbol that triggered the crash, at which point the
developer can annotate it for sharing.  In some cases, the crash can be a
genuine violation; e.g., a library exposes internal state to external
libraries, in which case the developer can decide to rework the library's API
to address the privacy issue.  This case is much less frequent and left at the
developer's discretion. An example is \texttt{ramfs}, which is so deeply
entangled with \texttt{vfscore} that blindly isolating it without redesign
would impair performance with little additional security benefits, as a
critical portion of the component's state would be shared. However, coupled
with \texttt{vfscore}, both components can perfectly well be isolated from the
rest of the system. This highlights a limitation of automated tools that
blindly isolate this component~\cite{Sartakov2021}.  Overall, the porting
process is greatly simplified by common debugging tools: GDB and all usual
debugging toolchains are supported.  The debugging experience in \sys is not
significantly different from Unikraft and most mainstream OSes, and we expect
it to remain intuitive for anyone familiar with OS development.  Depending on
the amount of data shared with the outside world, the porting process ranges
from 10 minutes (time subsystem, no data shared), to 2-5 days (filesystem, network
stack).  This porting cost is similar that of other compartmentalization
frameworks~\cite{Narayan2020}.  Table~\ref{tab:porting} illustrates the porting
effort with concrete numbers.

\subsection{Software Hardening}
The flexible isolation provided by \sys allows to enable/disable software
hardening (SH) such as CFI, etc., on a per-component basis:
isolating components without SH from components with it allows the latter to
maintain the guarantees offered by SH. Moreover, many SH schemes work by
instrumenting the memory allocator, and we use \sys' capacity to have an
allocator per-compartment to enable flexible SH. This flexibility allows for
example to alleviate the performance impact of SH by enabling it only for a
subset of the system. Our prototype currently uses address sanitization (KASan),
undefined behavior sanitization (UBSan), CFI, and stack protector.

%For example, in one scenario, using an
%instrumented allocator for hardened libraries and a non instrumented one for
%the rest, we measure a 14\% improvement in performance.

%\subsection{\lang Parser and Solver}
%We uses the parser Lark~\cite{LARK} for
%parsing the specification written in \lang. The toolkit begins by parsing the
%metadata of each of the application's component iteratively. After we create
%the initial component graph and encode the properties, we feed all the
%specifications to the Z3 Theorem Prover~\cite{Z3}. We then add the
%constraints and the order of relations between properties. Armed with all the
%logic, the solver returns the set of conforming \sys configurations.
%The solver also indicates notable configurations such as those with the least
%number of compartments and those with the least number of cross compartment
%edges. Finally, it outputs information about the degree of connectivity between
%components, which can be useful to understand performance during the design
%space exploration and to drive the said exploration.

%% file: 05-exploration.tex
\begin{figure}
    \center
    \includegraphics[width=0.49\textwidth]{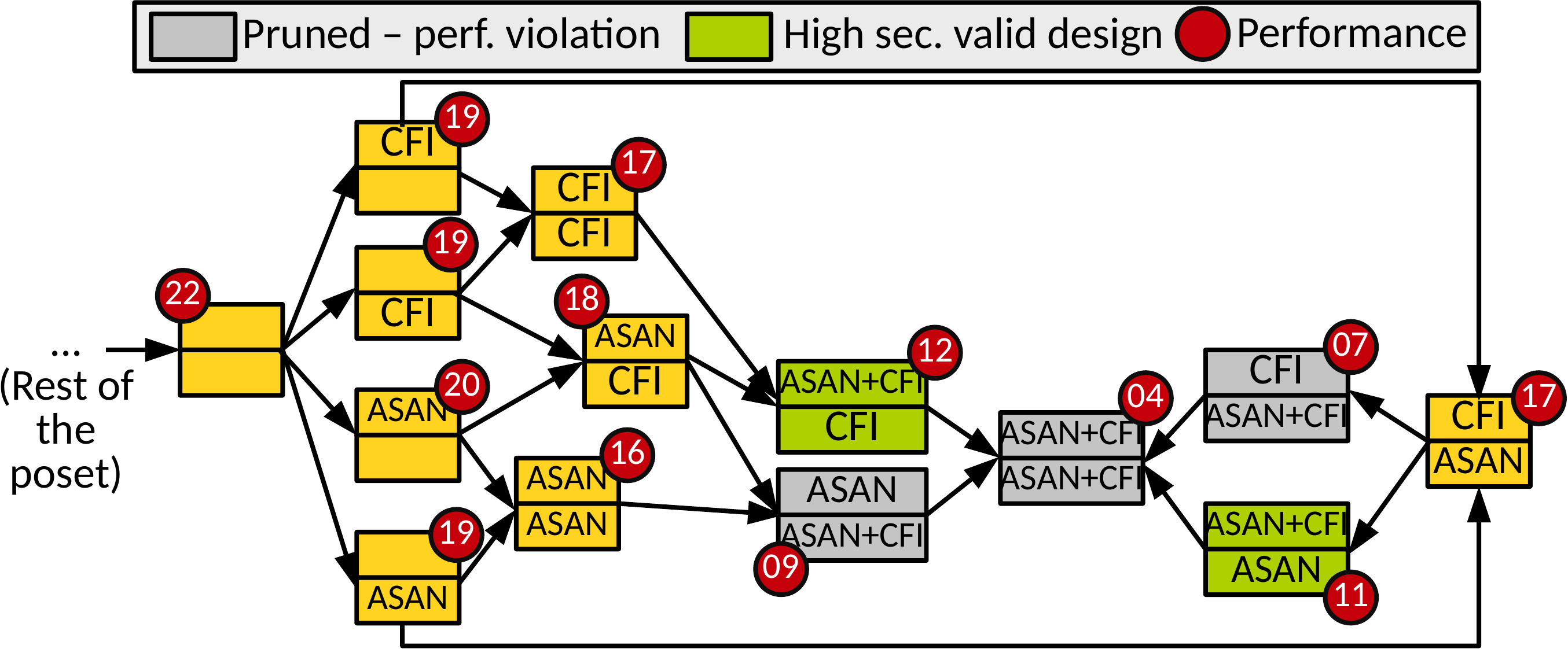}
    \caption{Partial view of the configuration poset for a fixed
        compartmentalization (2 compartments), varying per-compartment software
        hardening (CFI/ASAN). }
    \label{fig:exploration}
\end{figure}

\section{Exploration with Partial Safety Ordering}\label{sec:exploration}

In this section we present a design space exploration technique, \emph{partial
safety ordering}, that aims to guide a user towards suitable configurations for
a given use case by subsetting the vast design space enabled by \sys according
to safety and performance requirements.

% Needed for good figure placement, not sure how to do it otherwise...
\begin{figure*}
    \center
    \includegraphics[width=1.0\textwidth]{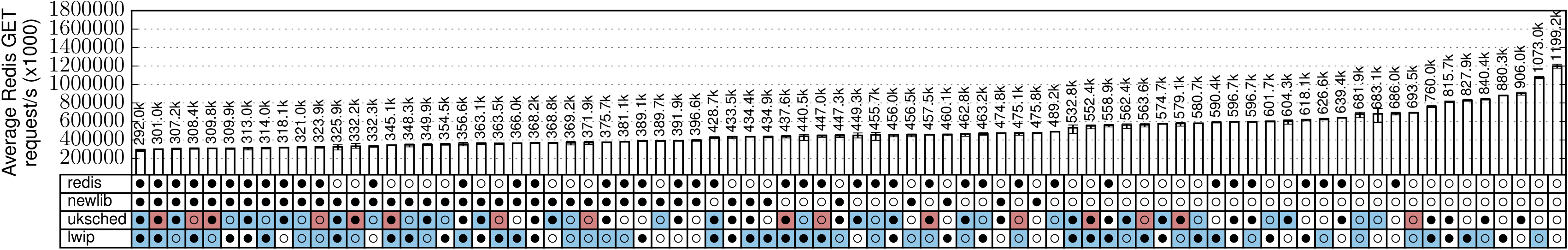}
    \includegraphics[width=1.0\textwidth]{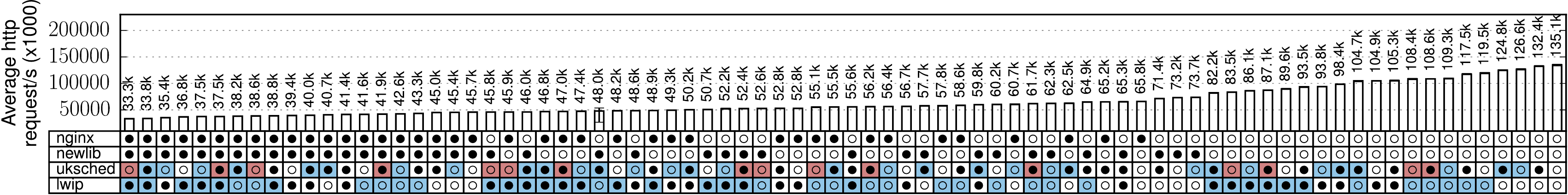}
    \caption{Redis (top) and Nginx (bottom) performance for a range of
        configurations. Components are on the left. Software hardening can be
        enabled [$\bullet$] or disabled [$\circ$] for each component. The
        white/blue/red color indicates the compartment the component is placed
        into. Isolation is achieved with MPK and DSS.}

    \label{fig:redis}
\end{figure*}

%\subsection{Partial Safety Ordering}
\label{subsec:exploration}

Given a performance budget, partial safety ordering attempts to
find the most secure configurations among those enabled by \sys. Quantifying
safety is challenging: it is impossible to give each configuration
an absolute safety score that would allow to completely order them; for
instance, is the safety of a configuration with 3 compartments, MPK isolation
and no hardening better or worse than another one with 2 compartments, EPT
isolation and CFI hardening?

Nevertheless, the safety of \emph{some} configurations is programmatically
comparable. Consider 3 configurations, \emph{C1} with no isolation and no
software hardening; \emph{C2} with two compartments protected by a given
mechanism with a given data sharing strategy and no hardening; and \emph{C3}
adding CFI for each compartment on top of \emph{C2}. In terms of
(probabilistic) safety, we have the following relationship: $C1 \leq C2 \leq
C3$. With that in mind, it is thus possible to organize all configurations into
a \emph{partially ordered set} (poset), that can be viewed as a Directed
Acyclic Graph (DAG) for which each node represents a configuration, and a
directed edge between nodes \emph{n1} and \emph{n2} indicates that the level of
safety of \emph{n1} is probabilistically superior to that of \emph{n2}. The
safety of nodes on the same path is comparable, while that of nodes on
different paths is not.

Figure~\ref{fig:exploration} presents a subset of the
configuration poset corresponding to fixed choices for a compartmentalization
strategy with 2 compartments, an isolation mechanism, and a strategy of data
sharing. This subset of the poset represents the variation of the last feature,
the software hardening, for which we assume only CFI and ASAN
for the sake of simplicity. Each configuration is depicted by a node
indicating, for each of the two compartments, which hardening mechanism is
applied: none, CFI, ASAN, and CFI+ASAN. We construct the poset partially depicted
on Figure~\ref{fig:exploration}, ordering safety with the assumption that
safety probabilistically increases with 1) the number of compartments; 2) data
isolation (isolated vs. shared stacks, dedicated shared memory areas per pair
of communicating compartments vs. shared areas accessible from everywhere,
etc.); 3) stackable software hardening; and 4) the strength of the isolation
mechanism.

Given such a poset, we can label each node with its performance characteristics
(circles in the figure denote fictional performance numbers), and prune those
that don't meet minimum requirements (gray nodes), ultimately yielding a set of
configurations that offer the best guarantees for a given performance budget.
This set corresponds to the \emph{maximal elements} of the poset, i.e. sinks of
the DAG (green nodes in the figure).

\paragraph{Partial Safety Ordering in Practice.}

In practice, users provide the toolchain with a test script (e.g., \texttt{wrk}
for Nginx) and a performance budget (e.g., at least 500k req./s).  Users are
free to define performance as they may deem suitable depending on their needs:
application throughput, tail latency, runtime, etc. Any metric is suitable as
long as it remains comparable across configurations and runs.  With this in
hand, the toolchain generates the unlabeled poset. Then, it labels it by
automatically measuring the performance of each configuration. The toolchain
does not have to run all configurations: assuming monotonically decreasing
performance, it can safely stop evaluating a path as soon as a threshold is
reached.  In practice, we observe that this significantly limits combinatorial
explosion. The result is a set of the most secure configurations for the given
budget, which the user can use to choose the most suitable one for a given use
case.  Ultimately, we expect this process to significantly trim the design
space and allow the user to make an informed and relatively effortless choice.

This approach assumes that the user is able to get representative feedback on
the application's performance, and users will not be able to use FlexOS’
exploration facilities if they are not able to properly benchmark their
application. However, we expect this situation to be quite rare: in the vast
majority of cases, users will be able to at least minimally test their
applications. These results can be used to exclude configurations that are too
costly and test the best candidates in production using lightweight performance
measurement systems, e.g., blue-green deployments.

\paragraph{Skipping Exploration.}

Some developers might already come with a particular isolation strategy in
mind. In that case the developer can skip this exploration phase by providing a
configuration file as shown in Section~\ref{sec:design}. In this case, the
developer leverages FlexOS’ flexibility and not its exploration facilities. We
note, however, that this ``expert'' approach has its limits: applications evolve
over time and a compartmentalization approach that is deemed optimal at a given
time may not be suitable in the future~\cite{Gudka2015}. In this case, an
exploration system such as FlexOS’ can be of use for the expert to easily
reconsider their approach in light of changing software.

%% file: 06-evaluation.tex
\section{Evaluation}\label{sec:evaluation}

We aim to demonstrate the vast performance/safety design space enabled by \sys,
assess the efficiency of the partial safety ordering exploration technique, and
compare \sys' performance with the literature. To this end, we present an
overview of the performance obtained with numerous safety configurations on
three popular cloud applications (Redis, Nginx, and SQLite), as well as iPerf,
a standard network stack benchmark. We demonstrate our design-space exploration
technique with Redis and Nginx. Then, we compare selected SQLite configurations
with Linux, CubicleOS~\cite{Sartakov2021}, a (non-flexible) compartmentalized
\LibOS, the SeL4~\cite{Klein2009}/Genode~\cite{Feske2021} microkernel, as
well as Unikraft~\cite{Kuenzer2021}.  Finally, we study raw isolation overheads in
\sys: DSS efficiency and cross-compartments call gate latencies.

We run experiments on an Intel Xeon Silver 4114 @2.2 GHz. For each experiment
we use 4 cores from the same socket, isolated with \emph{isolcpu}: 2 cores for
the client (iPerf client/\texttt{redis-benchmark}, \texttt{wrk}) on the host, 1
core for the QEMU process, and 1 core per \sys' vCPU. Hyperthreading is
disabled.

\subsection{Design Space Exploration: Redis, Nginx}

We automatically generate and run a large set of configurations for Redis and
Nginx using the Wayfinder~\cite{WAYFINDER} benchmarking platform.  We fix the
isolation mechanism to MPK with DSS and vary: the number of compartments (1-3),
compartmentalized components (TCP/IP stack, libc, scheduler, application), as
well as per-compartment software hardening (stack protector, UBSan and KASan),
for a total of 2x80 configurations.

\paragraph{Redis.}

The results are on Figure~\ref{fig:redis} (top), plotting for each
configuration Redis' GET throughput. Overall we observe that \sys enables for a
very wide range of safety configurations with significant performance
variation: there is one order of magnitude of difference between the
configuration yielding the lowest throughput (292K~req/s) vs. the highest one
(1.2M req/s).

Unsurprisingly, the configuration that disables isolation and hardening gives the
highest throughput. Conversely, configurations with many compartments/hardening
perform worst. Still, in between these two extremes,
creating more compartments and enabling hardening has a variable impact on
performance.  For example, with two compartments and no hardening, isolating
LwIP from the rest of the system leads to an 11\% performance hit, while that
number reaches more than 43\% when the scheduler is the isolated component ---
indicating extensive communication between user code and the scheduler. The
same is true for hardening: with a single compartment, enabling hardening on
the scheduler has a 24\% performance cost, while that cost is 42\% when
hardening the Redis application code.

The complexity of maximizing safety and performance becomes more clear when
isolating several components: isolating LwIP from the scheduler
from the rest only differs from a few percentage points from isolating LwIP
together with the scheduler from the rest. Such ``isolation for free'' effects
are caused by communication patterns; LwIP does not directly communicate with
the scheduler, hence the ``cut'' is not on a hot path, and merging them in a
same compartment brings little performance benefits.
Thus, the performance does not entirely depend on the number of compartments or
the number of components with hardening enabled, but rather \emph{what}
particular components are isolated/hardened, and their communication patterns.
Such effects can be leveraged to maximize safety and performance.

%\paragraph{Safety Constraints Solving.}
%\label{sec:scs-eval}

%We defined 5 Trust Models (TM) for Redis, and wrote the \lang's specifications
%for each. In TM0 the kernel components (lwip/sched) requirements make that they
%need to be isolated from the user code (Redis/newlib), for example in the case
%of verified kernel code/unsafe user code. TM1 requires to further split the
%kernel into two components, for example in the case of a verified scheduler, a
%mostly memory safe (e.g. Rust) TCP stack, and unsafe user code. TM2/TM3
%indicate that lwip/sched can be merged with user code, for example for
%performance reasons. Finally, TM4 assumes that the entire system is trusted.
%The conforming configurations that our solver outputs are illustrated on the
%last row of Figure~\ref{fig:redis}.

%In Figure~\ref{fig:redis} we have 4 trust models.
%A trust model is a compartmentalization scheme outputed by the solver given 
%the metadata of the components. For example, in this figure trust model 1 is the 
%case where the properties of all components hold if the scheduler and lwip are in
%separate compartments. Deriving configurations by applying SH are part of the same
%trust model. We can also move between trust model with a configuration given that
%SH changes the properties of a library such that it can be merged in another
%compartment(e.g. we apply SH on the scheduler and merge it with the rest of the kernel,
%thus moving to trust model 3).

\paragraph{Nginx.}

\begin{figure}
    \center
    \includegraphics[width=0.45\textwidth]{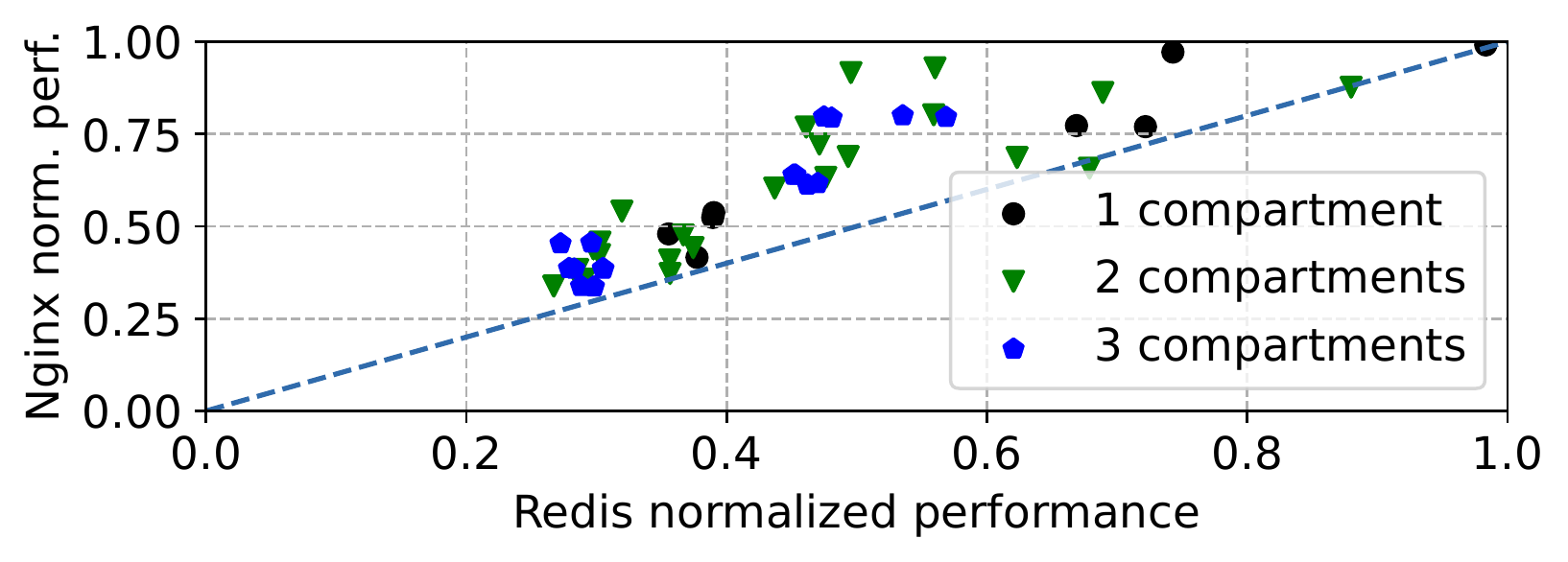}
    \caption{Nginx versus Redis normalized performance.}
    \label{fig:nginx-redis-scatter}
\end{figure}

The results are on Figure~\ref{fig:redis} (bottom), plotting for each
configuration Nginx' HTTP throughput. Overall we observe that results span over
the same range of overhead as Redis (0-4.1x).  However, overheads do not follow
the same distribution: 9 configurations have less than 20\% overhead in the
Nginx case, but only 2 for Redis. Similarly, 32 configurations have less
than 45\% of overhead, only 20 for Redis. This can be explained by looking more
closely at individual configurations.  Compared to Redis, isolating the
scheduler is much less expensive (6\% versus 43\% for Redis), and the same goes
for hardening (2\% versus 24\% for Redis). The costs, however, become similar
as more hardening and isolation boundaries are added because of bottleneck
effects.

This different distribution of costs is made more clear by
Figure~\ref{fig:nginx-redis-scatter} which compares the relative performance
of configurations for Nginx and Redis (same dataset as Figure~\ref{fig:redis}).
These differences show that isolating and hardening the \textit{same
  components} on two networked applications results in uneven, difficult to predict slow-down.
Existing approaches assume a one-size fits all safety configuration are therefore
suboptimal; in contrast, \sys enables users to easily navigate the safety / performance trade-off
inherent in their application.

\subsection{Partial Safety Ordering}

\begin{figure}
    \center
    \includegraphics[width=0.45\textwidth]{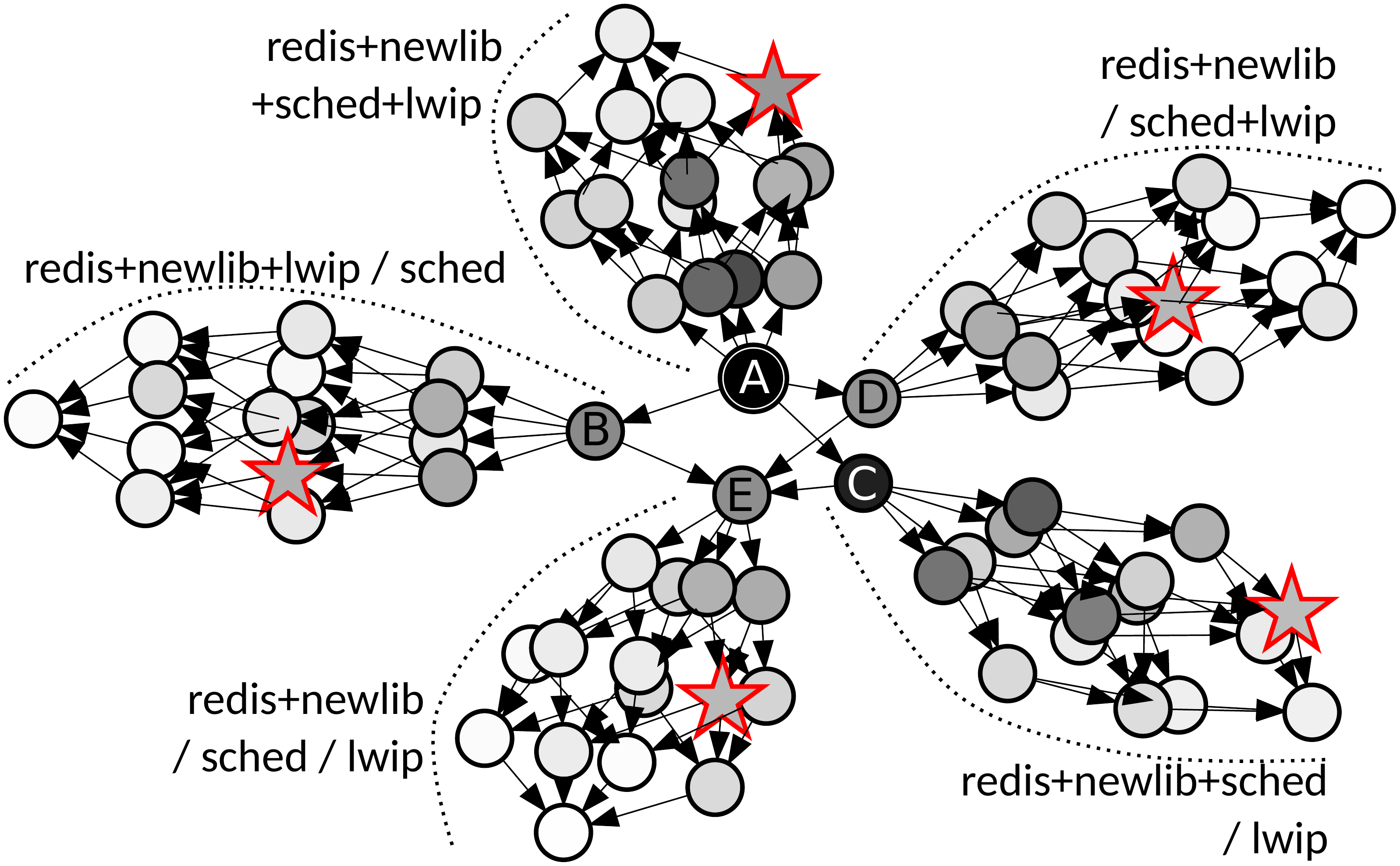}
    \caption{Configurations poset for the Redis numbers
    (Figure~\ref{fig:redis}). Stars are the most secure
    configurations with performance >= 500k requests/s.}

    \label{fig:redis-poset}
\end{figure}

We applied this technique on the Redis numbers from Figure~\ref{fig:redis}. We
construct the poset presented in Figure~\ref{fig:redis-poset}, where each node
is a Redis configuration, i.e. a column from Figure~\ref{fig:redis}. The node's
color intensity indicates the configuration's performance, black being the
fastest (1.2M~req/s) and slower configurations becoming gradually white (pure
white representing 292K~req/s). The fastest configuration is the one with
no isolation and no hardening (\BC{A} on
Figure~\ref{fig:redis-poset}). Other nodes in the center of the plot represent
compartments addition, still with no hardening: separating from the rest of the
system either the scheduler~\BC{B}, lwip~\BC{C}, or Redis+newlib~\BC{D}, and a
3 compartments scenario~\BC{E}. From these 5 basic compartmentalization
strategies come out 5 ``branches''. The nodes in each branch represent various
combinations of per-component software hardening. The nodes' color evolution
indicate the variable performance impact of creating new compartments and
stacking software hardening on components.

We set a minimum required performance of 500K~req/s, and let partial safety
ordering identify the safest configurations satisfying that constraint,
indicated with stars on Figure~\ref{fig:redis-poset}. In this case, the
technique can prune the configuration space from 80 to 5 configurations,
helping the user easily pick the most appropriate one.

\subsection{Batching Effects: Network Stack Throughput}

We port a simple iPerf server to \sys and use it to measure the network
performance of our system. We fix the compartmentalization to the following
scenario: the iPerf application code is placed within a compartment, and the
rest of the system (including the network stack) is placed in a second
compartment. We apply no software hardening, and configure the iPerf server to
pass buffers of varying sizes when calling \texttt{recv} on the socket. We measure the achieved throughput using an iPerf
client for \sys without isolation, with MPK (sharing or protecting the call
stack), as well as EPT. We run vanilla Unikraft as baseline.

\begin{figure}
    \center
    \includegraphics[width=\linewidth]{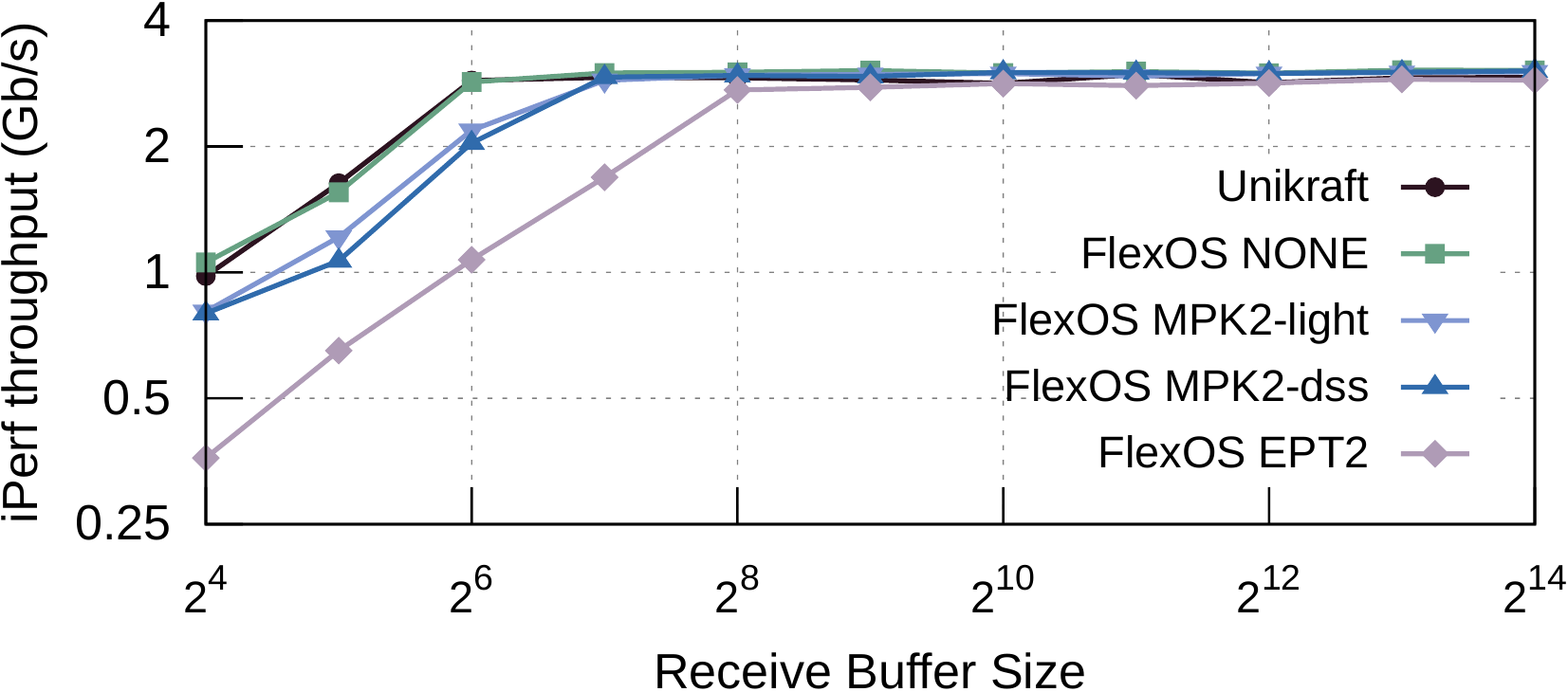}
    \caption{
Network stack throughput (iPerf) with Unikraft (baseline), \sys without
isolation, with two compartments backed by MPK (\textit{-light} = shared call
stacks, \textit{-dss} = protected and DSS), and with two compartments backed by
EPT.
}
    \label{fig:iperf}
\end{figure}

The results are on Figure~\ref{fig:iperf}. \sys without isolation performs
similarly to Unikraft, confirming that users ``only pay for what they get''.
\sys' isolation slowdown manifests for small payload sizes, for which the
domain crossing latency is an important bottleneck in the request processing
time. Depending on the buffer size, EPT isolation is 1.1-2.2x slower than MPK
with DSS, which is itself 0-1.5x slower than the baseline without isolation.
MPK with shared stacks bears a 0-1.3x slowdown.  Although MPK with DSS pays the
price of a stack switch (see Table~\ref{fig:gate-latency}), it is more secure
than fully sharing the stack and still faster than fully isolating it while
moving shared data to the heap (see Figure~\ref{fig:dss}). Batching effects
clearly manifest as the payload size increases: MPK's performance quickly
becomes similar to to baseline starting from 128~B.  EPT's isolation being more
costly, the payload size needs to be 256~B or above so that its performance to
reach about 90\% of the baseline's. These results illustrate that, depending on
the size of the payload and the frequency of domain crossings, all backends can
constitute a valid solution to a given problem.

\subsection{Filesystem Intensive Workloads: SQLite}

\begin{figure}
    \center
    \includegraphics[width=\linewidth]{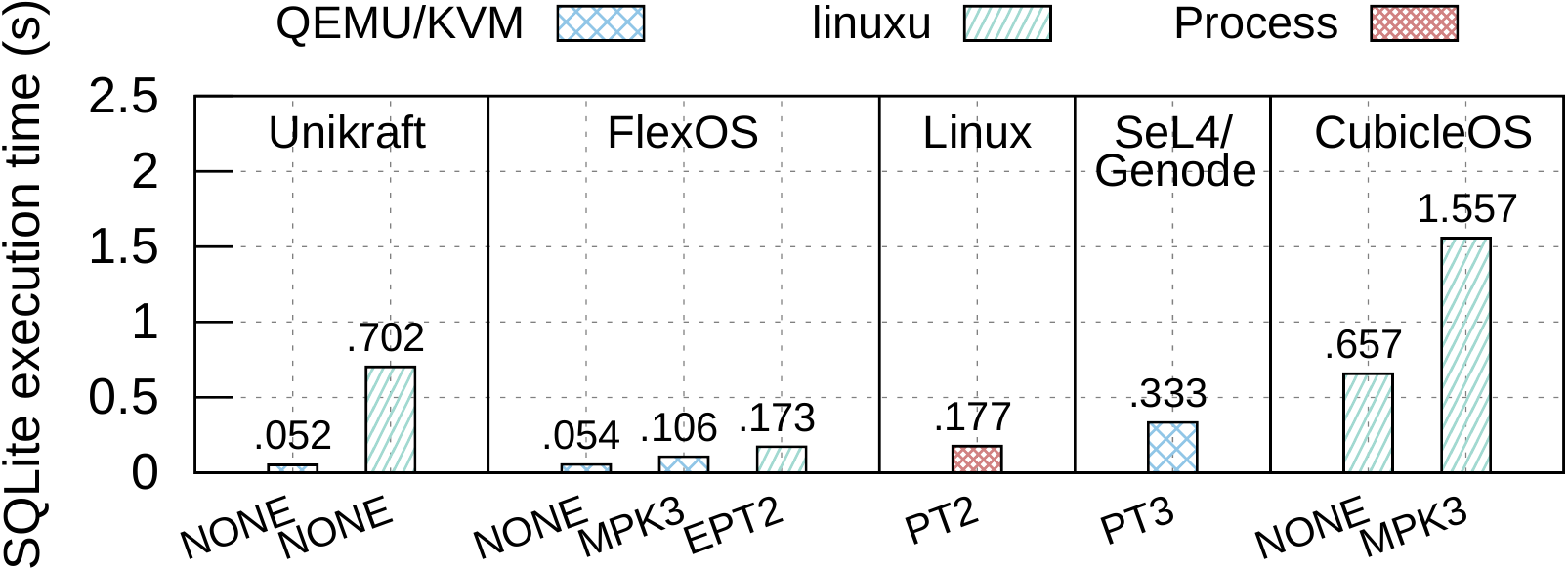}
    \caption{
Time to perform 5000 INSERT queries with SQLite on Unikraft, \sys, Linux, SeL4
(with the Genode system), and CubicleOS. The isolation profile is shown on the
x axis (NONE: no isolation, MPK3: MPK with three compartments, EPT2: two
compartments with EPT, PT2/3: two/three compartments with page-table-based
isolation).
}

    \label{fig:sqlite}
\end{figure}

We evaluate the performance of \sys with filesystem intensive workloads and
compare it to vanilla Unikraft, Linux, SeL4~\cite{Klein2009} with the
Genode~\cite{Feske2021} system, and CubicleOS~\cite{Sartakov2021}. Although
both \sys and CubicleOS extend Unikraft, the former runs in a standard Qemu/KVM
VM while the latter is implemented on top of \emph{linuxu}, Unikraft's Linux
userland debug platform. The Unikraft baseline number thus cover both cases.
We evaluate two scenarios: one with two components (EPT2, PT2), where the
filesystem is isolated from the application, and one with three components
(MPK3, PT3), where the filesystem is isolated from the time subsystem from the
rest of the system.  This benchmark performs 5000 INSERTs queries sequentially.
To increase pressure on the filesystem, each query is in a separate
transaction.  The results are shown in Figure~\ref{fig:sqlite}.

Compared to the baseline, \sys without isolation adds no overhead, and MPK3
adds an overhead of 2x.  This is still significantly faster than the userland
Linux version which performs a large number of system calls, highlighting the
benefits of the \LibOS basis.  Somewhat surprisingly, \sys with EPT2 performs
almost identically to Linux. This is because the syscall latency is almost
identical to the EPT2 gate latency on this system (see
Figure~\ref{fig:gate-latency}). Compared to SeL4, \sys is 3.1x faster with
MPK3, and 2x faster with EPT2.

Compared to CubicleOS, \sys is an order of magnitude faster. This is due to (1)
CubicleOS relying on \textit{linuxu}, i.e. running in Ring 3 and performing
Linux system calls for privileged operations, (2) CubicleOS not implementing
MPK support and relying on Linux \texttt{pkey\_mprotect} system calls (making
domain transitions orders of magnitude more expensive and the TCB thousands of
times larger), and (3) CubicleOS' \textit{trap-and-map} approach (that \sys
avoids with shared data annotations).  Even compared to its baseline without
isolation, CubicleOS with MPK3 adds an overhead of 2.4x, about 30\% more than
\sys.  CubicleOS without isolation is faster than the Unikraft linuxu baseline;
this is because it uses the Lea~\cite{Lea1996} memory allocator which behaves
better than Unikraft's TLSF~\cite{Masmano2004} allocator in this benchmark.

%\subsection{Micro-Benchmarks}
\subsection{Overheads: Stack Allocations, Gate Latencies}
\label{subsec:microbench}

% \paragraph{Component rewriting.}

% Changing the properties of a component can be achieved in multiple ways, among
% them rewriting in a different programming language, applying hardening mechanism
% and isolating it. In Figure~\ref{fig:rewriting} we measure the performace of a
% small http server that has a component that does the parsing. We note that each
% method requires a different degree of work, from close to no work when using 
% software hardening to a high amount of professional work to formally verify
% the component in Dafny.

% \begin{figure}
%     \center
%     \includegraphics[scale=.25]{include/rewriting}
%     \caption{Performance of a \sys simple http server when we rewrite and apply
%     different mechanisms on the parser component}
%     \label{fig:rewriting}
% \end{figure}

In \sys, stack data can be shared via heap allocations, using the DSS (trading
space for performance), or sharing the stack entirely (trading safety for
performance). To illustrate the benefits of the DSS, we measure, for each of
the mechanisms, the execution time of a function that allocates 1 to 3 shared
stack variables (size 1 Byte) and returns immediately.

\begin{figure}[]
\centering
\subfloat[Allocation latencies]{
    \includegraphics[width=.22\textwidth]{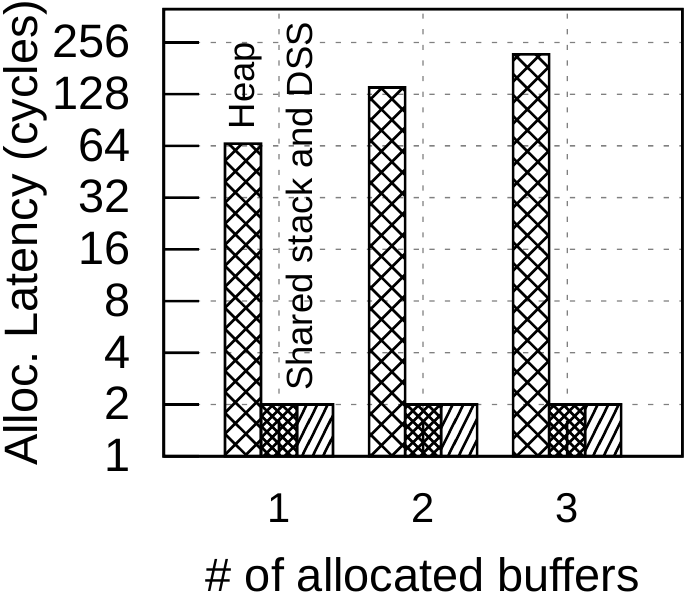}
    \label{fig:dss}
}
\subfloat[Gate latencies]{
    \includegraphics[width=.22\textwidth]{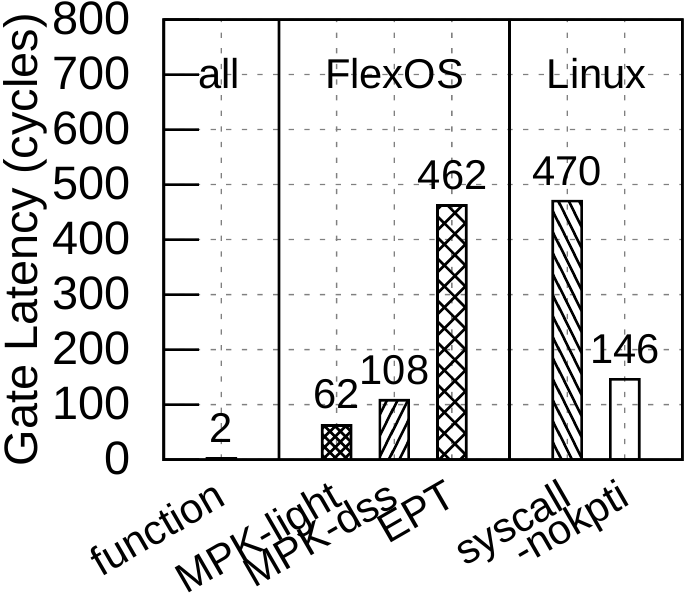}
    \label{fig:gate-latency}
}
\caption{\sys latency microbenchmarks.}
\end{figure}

The results are on Figure~\ref{fig:dss}. Heap-based stack allocations are one
to two orders of magnitude (100-300+ cycles) slower than typical stack
allocations (constant 2 cycles). This is not surprising, since general-purpose
allocators typically feature unbounded execution time.  This cost increases
with the number of variables, since each variable triggers a separate call to
malloc. The DSS matches the shared stack in performance, confirming that it
combines the safety of isolation with the performance of traditional stack
allocations. The memory footprint increase due to the DSS is modest as \sys
uses small stacks (8 pages). For example, an instance with Redis (8 threads),
has a space overhead of 288 KB. The DSS is a data sharing strategy and does not
remove the need to perform stack switches.

%\begin{table}
    %\center
    %\includegraphics[width=0.45\textwidth]{include/table-gate-latency}
    %\caption{\sys' gates latencies.}
    %\label{tab:gate-latency}
%\end{table}

Another source of compartmentalization overhead is gate latency. To illustrate
the raw performance of \sys' gates we measure the gate latency of MPK
stack-sharing gates (\textit{-light}), normal MPK gates, and EPT gates. We
compare them with the latency of a function call, and of a Linux system call
(with and without KPTI, \textit{-nopkti}).  The results are shown in
Figure~\ref{fig:gate-latency}. MPK light gates are 80\% faster than normal MPK
gates, and 7.6x faster than EPT gates, as they correspond to the cost of raw
\texttt{wrprku} instructions. EPT latencies are similar to syscall latencies
without KPTI, illustrating the practicability of the EPT backend.

%% file: 06bis-impact.tex
\section{Use Cases for Isolation Flexibility}

\sys enables developers to seamlessly experiment with various safety
configurations for their OS. An obvious use-case we presented throughout this
paper is the specialization of the OS' safety strategy for a given application:
manually or semi-automatically selecting, among the vast design space unlocked
by \sys, the most suitable configuration for a particular use case with given
safety/performance constraints. Still, there are many other ways in which this
flexibility can be used; we detail some of them next.

\paragraph{Quickly Isolate Exploitable Libraries.} Consider the period between
the full disclosure of a vulnerability and the release of its fix, or the
embargo period when vulnerabilities are disclosed only to affected vendors, but
not to the general public; these periods can last for weeks up to years during
which vulnerable software runs in the wild.  With \sys, it takes seconds to
create a new binary that isolates a vulnerable library into its own compartment
(e.g. EPT + hardening) to at least mitigate the effects of exploits; an
automated system could be created to respond to known vulnerabilities by
recompiling production software to isolate certain libraries, similar to
Self-Certifying Alerts~\cite{sca}. Such flexibility improves over the state of
the art by avoiding a loss of functionality (e.g. compared to
Senx~\cite{Huang2019}), and providing excellent resistance to polymorphic
variations of vulnerabilities (e.g. compared to filters~\cite{Costa2007}).

\paragraph{Quickly React to Hardware Protections Breaking Down.} Recent
hardware vulnerabilities~\cite{MELTDOWN, SPECTRE} showed that hardware-backed
isolation mechanisms are not foolproof. The corresponding fixes may require
significant engineering and redesign efforts (e.g. KPTI for Meltdown), leading
to long vulnerability windows. \sys is not immune to hardware vulnerabilities
by design. In this case, however, its ability to easily switch between
protection techniques comes handy: by supporting a wide range of isolation
primitives relying on a range of different hardware, switching the isolation
mechanism from a vulnerable to a non-vulnerable one is just a matter of
rebuilding the LibOS with a different configuration (snippet in
\S\ref{sec:design}), i.e. the engineering cost is nil.

\paragraph{As Secure as You can Afford.} Consider a service provider who wishes
to offer the best possible security as long as its server can keep up with the
client load. A natural approach would be to run the safest combination that
copes with peak load, as we suggested in our Redis evaluation; this means that
in periods of low load the system has idle compute power.

With its capacity to quickly switch safety configurations, \sys enables another
approach: to run, at any time, the safest configuration that can sustain the
actual load. This makes attacks much harder as long as the system is
under-loaded, but gracefully switches off defenses as load increases to
respect SLA. Another approach is to couple this with software load balancers
to triage users into likely benign or malicious, sending them to machines
running faster or safer software, accordingly.

\paragraph{Dealing with Crashed Software.} Vulnerabilities are a fact of life,
and the standard approach is to quickly restart crashed software and to
examine the faults in the background. When such a crash is detected (e.g.
memory error), with \sys it is wiser to start a safer configuration of the same
software, to ensure that any vulnerability is not turned into an exploit.

\paragraph{Incremental Verification.} Individual components of \sys can be
verified and isolated from the rest of the system. In this way, one can
obtain strong guarantees on pre-conditions and ensure that verified properties
hold even when mixed with unverified components, something that isn't possible
with monolithic operations systems~\cite{Li2021}. Over time, the entire system
could be verified, gradually increasing the guarantees of the system.

\paragraph{Deployment to Heterogeneous Hardware.} The flexibility of \sys
mechanisms can also come in very handy when deploying on heterogeneous
hardware. Some servers might offer MPK support for example, others CHERI,
others only the classical MMU. In every case, Chrysalis is able to get the best
from the available hardware without major rewrite, and without requiring insider
knowledge from application developers.

%\paragraph{Early Design Space Exploration for a System under Construction.}
%When building an embedded system, the selection of the hardware/software
%composing it is made early in the design flow, but will define the system's
%performance, safety, power consumption, etc. Thus, system designers seek tools
%and methods to estimate as early as possible these metrics and control them
%in the final product. The flexibility of \sys, combined with the
%relatively low engineering cost of porting applications, enables a system
%designer to explore various OS safety strategies early in the design flow. As
%such, the impact of these strategies on metrics such as the performance/power
%consumption of the final product can be estimated.

%% file: 07-related-works.tex
\section{Related Work}\label{sec:related-works}

\paragraph{Improving OS safety.}

Previous work proposed to address the safety issues of monolithic OSes by
reducing the TCB through separation~\cite{Rushby1981, AlvesFoss2006},
micro-kernels~\cite{Golub1992, Herder2006}, and safe languages~\cite{Boos2020,
Narayanan2020a, Hunt2007, Madhavapeddy2013, Cutler2018}. In SASOSes, internal
isolation may be traded off for performance~\cite{Kantee2012, Kivity2014,
Olivier2019, Kuenzer2021}, provided with traditional page
tables~\cite{Chase1994, Leslie1996, Heiser1999, Nikolaev2020}, or intra-AS
hardware isolation mechanisms~\cite{Sartakov2021, Li2020, Sung2020,
Olivier2020}. Other research efforts strive to speedup IPC in
microkernels~\cite{Gu2020, Mi2019}, or redesign monolithic OSes
entirely~\cite{Haertig1997, Swift2002, Castro2009, LeVasseur2004,
BoydWickizer2010, Nikolaev2013,Dautenhahn2015}.

%Formal verification is ultimately the only way to offer deterministic safety
%guarantees, though it has trouble scaling to modern OSes' large
%codebases~\cite{Klein2009B, Klein2009}. Thus, several works focus on verifying
%only a subset of large OSes, such as drivers~\cite{Witkowski2007} or
%filesystems~\cite{Amani2016}. Runtime software hardening mechanisms with low
%overhead are commonly found in production kernels~\cite{Edge2013, Edge2014A}.
%However, the most safety-efficient ones~\cite{Edge2014B} are generally only
%enabled for safety test runs due to their high performance impact.

Overall, each of these approaches is a single or a few point(s) in the OS
safety/performance design space and lacks the flexibility of \sys to
automatically specialize for safety or performance.
LibrettOS~\cite{Nikolaev2020} allows a \LibOS to switch between SASOS and
microkernel modes, but remains limited to a small subset of the
safety/performance design space.

\paragraph{Compartmentalization Frameworks.}

Several compartmentalization frameworks have been proposed
recently~\cite{VahldiekOberwagner2019, Hedayati2019, Narayan2020,
Schrammel2020, Gudka2015, Liu2017, Bauer2021, Sartakov2021}. Contrary to \sys,
none focuses on flexible isolation. Regarding application porting,
most~\cite{VahldiekOberwagner2019, Hedayati2019, Narayan2020, Schrammel2020}
rely on code annotations. A few studies provide various degrees of porting
automation through data flow analysis~\cite{Gudka2015, Liu2017, Bauer2021}, but
are typically bound to numerous limitations due to the complexity of breaking
down monolithic code bases. Nevertheless, some of their principles can be
applied to increase the degree of automation of \sys' porting process --
something we scope out as future works. CubicleOS~\cite{Sartakov2021} proposes a
\textit{trap and map} mechanism to limit the porting effort, but this comes at
a high cost, is specific to MPK, and is not entirely automated. Further, as
shown in our evaluation, CubicleOS' reliance on Unikraft's \emph{linuxu} leads
to suboptimal performance.

%% file: 08-conclusion.tex
\section{Conclusion}\label{sec:conclusion}
The isolation strategy of today's OSes is mostly fixed at design time. This
lack of flexibility is problematic in many scenarios. We propose \sys, an OS
whose isolation strategy is decoupled from its design. We augment the
historical capacity of the \LibOS to specialize towards performance with the
ability to specialize for safety: fundamental decisions such as the
compartmentalization granularity and which isolation mechanism to use are
deferred to build time. \sys ships with a semi-automated exploration strategy
helping the user navigate the vast configuration space the system unlocks.
\sys is available online at \url{https://project-flexos.github.io} under an
open source license.

In our future work, we intend to add more isolation backend implementations to
\sys including CHERI and SGX, as well as support for more software hardening techniques. 
Another direction of future work is to create a formal basis to help users navigate
the safety configuration space. This would enable, among others, embedding formally verified
components in \sys configurations while preserving their proven properties.

%% file: ae.tex
% LaTeX template for Artifact Evaluation V20201122
%
% Prepared by 
% * Grigori Fursin (cTuning foundation, France) 2014-2020
% * Bruce Childers (University of Pittsburgh, USA) 2014
%
% See examples of this Artifact Appendix in
%  * SC'17 paper: https://dl.acm.org/citation.cfm?id=3126948
%  * CGO'17 paper: https://www.cl.cam.ac.uk/~sa614/papers/Software-Prefetching-CGO2017.pdf
%  * ACM ReQuEST-ASPLOS'18 paper: https://dl.acm.org/citation.cfm?doid=3229762.3229763
%
% (C)opyright 2014-2020
%
% CC BY 4.0 license
%

%\documentclass{sigplanconf}

%\usepackage{hyperref}

%\begin{document}

%\special{papersize=8.5in,11in}

%%%%%%%%%%%%%%%%%%%%%%%%%%%%%%%%%%%%%%%%%%%%%%%%%%%%
% When adding this appendix to your paper, 
% please remove above part
%%%%%%%%%%%%%%%%%%%%%%%%%%%%%%%%%%%%%%%%%%%%%%%%%%%%

\appendix
\section{Artifact Appendix}

%%%%%%%%%%%%%%%%%%%%%%%%%%%%%%%%%%%%%%%%%%%%%%%%%%%%%%%%%%%%%%%%%%%%%
\subsection{Abstract}

% Briefly and informally describe your artifact including minimal hardware and
% software requirements, how it supports your paper, how it can be validated,
% and what is the expected result. It will be used to select appropriate
% reviewers. It will also help readers understand what was evaluated and how.

This artifact contains the source code of \sys, the proof-of-concept of our
flexible isolation approach, along with all scripts necessary to reproduce the
paper's measurements and plots. The goal of this artifact is to allow readers
to reproduce the paper's results, and build new research on top of \sys.

\subsection{Artifact Check-List (meta-information)}

% Together with the artifact abstract, this check-list will help us make sure
% that reviewers have appropriate competency and an access to the technology
% required to evaluate your artifact. It can also be used as meta information
% to find your artifacts in Digital Libraries.

{\small
\begin{itemize}
  \item {\bf Program: } the \sys library OS, benchmarked with standard application
	  benchmarks (\texttt{wrk} and \texttt{redis-benchmark}), a custom SQLite
	  benchmark, and custom microbenchmarks.
  \item {\bf Binary: } automatically built from source.
  \item {\bf Run-time environment:} GNU/Linux Debian 11 (Bullseye), with KVM and Docker.
	  Other dependencies are automatically installed.
  \item {\bf Hardware:} Intel® Xeon® Silver 4114 @ 2.20 GHz, or any machine with more than 8 cores that supports Intel MPK,
	  typically Intel® Xeon®
	  Scalable Processors starting with the Skylake generation. At least 128.0~GB of RAM.
  \item {\bf Metrics:} requests/s, Gb/s, queries/s, execution time, gate latencies.
  \item {\bf Output:} performance data, \sys images.
  \item {\bf Experiments:} \Cref{fig:redis,fig:nginx-redis-scatter,fig:iperf,fig:sqlite,fig:dss,fig:gate-latency} are reproducible automatically. \Cref{fig:redis-poset} is reproducible manually (it is only a graph). \Cref{tab:porting} is also reproducible manually.
  \item {\bf How much disk space required (approximately)?:} 100.0~GB
  \item {\bf How much time is needed to prepare workflow (approximately)?:} 6-12~Hours (\textit{automated}).
  \item {\bf How much time is needed to complete experiments (approximately)?:} 4-5~Hours (\textit{automated}), and up to 1.5~Hours (\textit{manual}).
  \item {\bf Publicly available?:} Yes.
  \item {\bf Code licenses (if publicly available)?: } BSD-3-clause.
  \item {\bf Workflow framework used?: } Wayfinder~\cite{WAYFINDER}, Docker, scripts.
  \item {\bf Archived (provide DOI)?: } \texttt{10.5281/zenodo.5748505}
\end{itemize}
}

%%%%%%%%%%%%%%%%%%%%%%%%%%%%%%%%%%%%%%%%%%%%%%%%%%%%%%%%%%%%%%%%%%%%%
\subsection{Description}

\subsubsection{How to access}

The latest version of the artifact can found on
GitHub\footnote{\url{https://github.com/project-flexos/asplos22-ae}}.
Alternatively, individual releases can be downloaded from our Zenodo
archive\footnote{\url{https://zenodo.org/record/5748505}}.  Note that the
artifact evaluation (AE) GitHub repository only contains part of the artifact,
namely scripts to reproduce this paper's experiments. The core of \sys,
libraries, and applications, are all available in the \texttt{project-flexos}
organization, as documented in the AE repository.

In order to precisely reproduce this paper's measurements, we gave ASPLOS'22
reviewers access to our server, an Intel® Xeon® Silver 4114 with 128.0 GB RAM,
Debian 11.1, and Linux version \texttt{5.10.70-1}. Nonetheless, access to this
particular setup is not required to run this artifact; hardware and software
dependencies are detailed further below.

\subsubsection{Hardware dependencies}

An Intel® Xeon® Silver 4114 @ 2.20 GHz, or any machine that supports Intel MPK,
typically any Intel® Xeon® Scalable Processor starting with the Skylake
generation. The processor must have more than 8 cores. 128.0~GB of RAM are
necessary to run the experiments corresponding to \Cref{fig:redis}, as all
images are built and stored in RAM by our tool in order to achieve reasonable
preparation times. Note that this amount of cores/RAM is required to reproduce
this paper's results, \textit{not} to run \sys.

\subsubsection{Software dependencies}
\label{ssec:deps}

This artifact has been tested with Debian GNU/Linux 11 (Bullseye) with Linux
kernel version \texttt{5.10.70-1} (KVM enabled), Docker version
\texttt{20.10.10} (or any recent version). All other dependencies are
automatically installed by the artifact's scripts.

\subsubsection{Data sets}

All data sets and benchmarks are included in the artifact, generated
automatically, or downloaded automatically by the run scripts.

%%%%%%%%%%%%%%%%%%%%%%%%%%%%%%%%%%%%%%%%%%%%%%%%%%%%%%%%%%%%%%%%%%%%%
\subsection{Installation}

Before running any experiment, prepare your host with the recommendations
detailed above in \ref{ssec:deps}. Note that all commands below assume
superuser permissions. Once the system is set up, clone our AE
repository:

\begin{tcolorbox}[colback=lightgrey,boxrule=0pt,arc=0pt,left=6pt]
{\scriptsize
\begin{Verbatim}[commandchars=\\\{\}]
$ git clone https://github.com/ukflexos/asplos22-ae.git
\end{Verbatim}
}
\end{tcolorbox}

Then, generate a GitHub personal access token with the permissions
"\texttt{public\_repo}" and set it in the Makefiles. You can do it for the
entire system by exporting an environment variable:

\begin{tcolorbox}[colback=lightgrey,boxrule=0pt,arc=0pt,left=6pt]
{\scriptsize
\begin{Verbatim}[commandchars=\\\{\}]
$ export KRAFT_TOKEN="<your token>"
\end{Verbatim}
}
\end{tcolorbox}

Alternatively, you can also set it individually in every Makefile by editing
the \texttt{KRAFT\_TOKEN} variable:

\begin{tcolorbox}[colback=lightgrey,boxrule=0pt,arc=0pt,left=6pt]
{\scriptsize
\begin{Verbatim}[commandchars=\\\{\}]
...
#
# Parameters
#
KRAFT_TOKEN ?= <your token>
...
\end{Verbatim}
}
\end{tcolorbox}

Note that if \texttt{KRAFT\_TOKEN} is set system-wide, definitions in Makefiles
will not override it. After this, install dependencies on the host:

\begin{tcolorbox}[colback=lightgrey,boxrule=0pt,arc=0pt,left=6pt]
{\scriptsize
\begin{Verbatim}[commandchars=\\\{\}]
$ make dependencies
\end{Verbatim}
}
\end{tcolorbox}

%%%%%%%%%%%%%%%%%%%%%%%%%%%%%%%%%%%%%%%%%%%%%%%%%%%%%%%%%%%%%%%%%%%%%
\subsection{Experiment Workflow}
\label{subsec:workflow}

All experiments should be prepared first. The prepare step installs necessary
tools and downloads additional resources before they can run. This can be done
for a single experiment or for all experiments, for example:

\begin{tcolorbox}[colback=lightgrey,boxrule=0pt,arc=0pt,left=6pt]
{\scriptsize
\begin{Verbatim}[commandchars=\\\{\}]
$ make prepare-fig-07 # prepare experiment 7
$ make prepare # prepare all experiments
\end{Verbatim}
}
\end{tcolorbox}

The automated preparation of all experiments takes on average 6-12~hours on our setup.
This very long preparation time is due to the generation of all images.
Once one or many experiments have been prepared they can be run, again using a
similar syntax as above:

\begin{tcolorbox}[colback=lightgrey,boxrule=0pt,arc=0pt,left=6pt]
{\scriptsize
\begin{Verbatim}[commandchars=\\\{\}]
$ make run-fig-07 # run experiment 7
$ make run # run all experiments
\end{Verbatim}
}
\end{tcolorbox}

Running all automated experiments takes on average 4-5~hours on our setup.  The
plot for \Cref{fig:redis-poset} is not automated, and neither is the
measurement of LoC changes for \Cref{tab:porting}. We estimate that the
combination of the two manual items may take up to 1.5~hours of manual work.

Automated experiments will generate the relevant experimental results within
the experimental folder of the specific experiment. To plot one or many
experiment figures, use, for example:

\begin{tcolorbox}[colback=lightgrey,boxrule=0pt,arc=0pt,left=6pt]
{\scriptsize
\begin{Verbatim}[commandchars=\\\{\}]
$ make plot-fig-07 # plot experiment 7
$ make plot # plot all experiments
\end{Verbatim}
}
\end{tcolorbox}

You can clean, or "properclean" to completely reset any preparation with \texttt{make
clean} or \texttt{make properclean} for individual or all experiments, for example:

\begin{tcolorbox}[colback=lightgrey,boxrule=0pt,arc=0pt,left=6pt]
{\scriptsize
\begin{Verbatim}[commandchars=\\\{\}]
$ make clean-fig-07
$ make properclean-fig-07
$ make clean
$ make properclean
\end{Verbatim}
}
\end{tcolorbox}

The clean rule removes results and plots, the properclean rule additionally
deletes containers.

%%%%%%%%%%%%%%%%%%%%%%%%%%%%%%%%%%%%%%%%%%%%%%%%%%%%%%%%%%%%%%%%%%%%%
\subsection{Evaluation and Expected Results}

Reproducing experiments on the same machine should produce the same results as
in the paper. On other machines, we expect different absolute numbers but
similar ordering. On recent processors that benefit from hardware mitigations
for transient executions attacks we expect EPT, Linux, and SeL4 measurements to
improve comparatively to the MPK baseline.

%%%%%%%%%%%%%%%%%%%%%%%%%%%%%%%%%%%%%%%%%%%%%%%%%%%%%%%%%%%%%%%%%%%%%
\subsection{Experiment Customization}

Reviewers may use the base \sys Docker container to access a clean \sys
development environment, port their own application, and build custom images.
Instructions to build the base FlexOS Docker image, port applications, and
build custom images are available in the \texttt{README.md} file of our main AE
repository\footnote{\url{https://github.com/project-flexos/asplos22-ae/blob/main/README.md}}.

%%%%%%%%%%%%%%%%%%%%%%%%%%%%%%%%%%%%%%%%%%%%%%%%%%%%%%%%%%%%%%%%%%%%%
\subsection{Notes}

Some experiments have a slightly different workflow compared to the one
described in \ref{subsec:workflow}. \Cref{fig:redis} requires you to set
\texttt{HOST\_CORES} with a set of cores to be used for the experiment.
\Cref{fig:nginx-redis-scatter} is only a plot and requires some manual steps.
\Cref{fig:gate-latency} requires a reboot of the machine with different kernel
parameters. \Cref{tab:porting} is manual. In all of these cases, the local
\texttt{README.md} provides appropriate explanations.  In general, the
top-level and individual \texttt{README.md} files of our artifact contains more
precise information on experiment timings, repository structure, setup
requirements, and potential issues and solutions.  We strongly recommend a
careful read of these instructions before starting to reproduce experiments.

%%%%%%%%%%%%%%%%%%%%%%%%%%%%%%%%%%%%%%%%%%%%%%%%%%%%%%%%%%%%%%%%%%%%%
\subsection{Methodology}

Submission, reviewing and badging methodology:

\begin{itemize}
  \item \url{https://acm.org/publications/policies/artifact-review-badging}
  \item \url{http://cTuning.org/ae/submission-20201122.html}
  \item \url{http://cTuning.org/ae/reviewing-20201122.html}
\end{itemize}

%%%%%%%%%%%%%%%%%%%%%%%%%%%%%%%%%%%%%%%%%%%%%%%%%%%%
% When adding this appendix to your paper, 
% please remove below part
%%%%%%%%%%%%%%%%%%%%%%%%%%%%%%%%%%%%%%%%%%%%%%%%%%%%

%\end{document}